\documentclass[sigconf]{acmart}
\acmSubmissionID{3072}

\ifdefined\iflatexml\else
\expandafter\newif\csname iflatexml\endcsname
\fi

\usepackage{color}
\usepackage{bm}
\usepackage{subcaption}
\usepackage{multirow}
\usepackage{enumitem}

\iflatexml
\newcommand{\compactheaderbreak}{\\}
\else
\makeatletter
\def\city#1{\global\@ACM@citypresenttrue\@ACM@addtoaddress{\mbox{#1}}}
\makeatother

\newcommand{\compactheaderbreak}{\\[-2pt]}
\fi

\copyrightyear{2026}
\acmYear{2026}
\iflatexml
\setcopyright{2026 Copyright held by the owner/author(s). Licensed under CC BY 4.0.}
\else
\setcopyright{cc}
\fi
\setcctype{by}
\acmConference[UIST '26]{The 39th Annual ACM Symposium on User Interface Software and Technology}{November 02--05, 2026}{Detroit, MI, USA}
\acmBooktitle{The 39th Annual ACM Symposium on User Interface Software and Technology (UIST '26), November 02--05, 2026, Detroit, MI, USA}
\acmDOI{10.1145/3830398.3830545}
\acmISBN{979-8-4007-2856-3/2026/11}

\providecommand{\added}[2][]{{#2}}
\providecommand{\deleted}[2][]{}

\newif\ifshowchanges
\showchangesfalse %

\definecolor{forestgreen}{RGB}{0, 140, 0}
\definecolor{cerulean}{RGB}{42, 82, 190}
\definecolor{blue}{RGB}{0, 0, 220}

\definecolor{changecolor}{RGB}{20, 50, 200}

\definecolor{abecolor}{RGB}{0, 140, 0}
\definecolor{abe_addcolor}{rgb}{0,0.3,0.8}
\definecolor{abe_removecolor}{rgb}{0.9,0.2,0.1}

\definecolor{nhancolor}{RGB}{0, 100, 50}

\ifshowchanges

    \newcommand{\adelete}[1]{{\leavevmode\textcolor{abe_removecolor}{{DeletedPart}}}}
    
    \newcommand{\acomment}[1]{\comment[id={\AbeD}]{#1}}
    \newcommand{\changenote}[2][]{\added[comment={#2}]{}}
    
    \newcommand{\nchange}[1]{\textcolor{nhancolor}{#1}}   %
    \newenvironment{revised}{
    \color{changecolor}
    }{
    \color{black}
    }

    \newenvironment{nhan_revised}{
    \color{nhancolor}
    }{
    \color{black}
    }

\else

    \newcommand{\adelete}[1]{}
    
    \newcommand{\acomment}[1]{}
    \newcommand{\changenote}[2][]{}
    
    \newcommand{\nchange}[1]{#1}

    \newenvironment{nhan_revised}{
    }{
    }

\fi

\newif\ifsubmit
\submittrue

\ifsubmit
\definecolor{author_colorA}{rgb}{0,0.5,1}
\definecolor{author_colorB}{rgb}{0.2,.64,0}
\definecolor{author_colorC}{rgb}{1,0,1}
\definecolor{author_colorD}{rgb}{0,1,1}
\definecolor{changes_color}{rgb}{0.05,0.5,0.3}
\definecolor{mathbrace_color}{rgb}{0.2,0.5,1.0}
\fi

\definecolor{abetextcolor}{rgb}{0,0.1,1.0}

\ifsubmit
    \newcommand{\mackenzie}[1]{}
    \newcommand{\valentina}[1]{}
    \newcommand{\abe}[1]{}
    \newcommand{\nhan}[1]{}

    \newcommand{\URGENT}[1]{}
\else
    \newcommand{\mackenzie}[1]{{\textcolor{orange}{[{ML}: #1]}}}
    \newcommand{\valentina}[1]{\textcolor{magenta}{[{VS}: #1]}}
    \newcommand{\abe}[1]{\textcolor{abetextcolor}{\textbf{{AD}: #1}}}
    \newcommand{\nhan}[1]{\textcolor{nhancolor}{#1}}

    \newcommand{\URGENT}[1]{{\textcolor{red}{URGENT:[#1]}}}
\fi

\newenvironment{packed_item}{
\begin{itemize}[leftmargin=0.35cm]
  \setlength{\itemsep}{1pt}
  \setlength{\parskip}{2pt}
  \setlength{\parsep}{0pt}
}{\end{itemize}}

\newcommand{\beq}{\begin{equation}}
\newcommand{\eeq}{\end{equation}}

\newcommand{\baeq}{\beq\begin{aligned}}
\newcommand{\eaeq}{\end{aligned}\eeq}

\newcommand{\defineterm}[1]{\emph{#1}}

\newcommand{\namedparagraph}[1]{\vspace{0.2cm}\noindent\textbf{#1:}}

\newcommand{\Sec}[1]{Section {#1}}

\newcommand{\Eqn}[1]{Equation {#1}}
\newcommand{\EqRef}[1]{\Eqn{\ref{#1}}}

\newcommand{\synchrony}{synchrony}
\newcommand{\synchronyfunc}{\sigma}

\newcommand{\nclips}{N}

\newcommand{\nEvents}{|\NarrativeSet{}|}

\newcommand{\SourceMediaSymbol}{M}

\newcommand{\narrative}{narrative}

\newcommand{\eventin}{\subset}

\newcommand{\roughmap}{\mathbf{r}}
\newcommand{\RoughCutSymbol}{R}
\newcommand{\RoughCut}{\mathbf{\RoughCutSymbol}}

\newcommand{\RoughCutClipi}[1]{\MediaMi{\roughmap{}(#1)}(\NElementi{#1})}

\newcommand{\element}{element}

\newcommand{\sourcemedia}{source media}
\newcommand{\SourceMediaSet}{\mathbf{\SourceMediaSymbol{}}}

\newcommand{\NarrativeSymbol}{S}
\newcommand{\elementinstancesymbol}{s}
\newcommand{\NarrativeSet}{\mathbf{\NarrativeSymbol}}
\newcommand{\NElementi}[1]{\elementinstancesymbol_{#1}}

\newcommand{\EventSetSymbol}{E}
\newcommand{\eventinstancesymbol}{e}
\newcommand{\EventSet}{\mathbf{\EventSetSymbol}}
\newcommand{\eventi}[1]{\eventinstancesymbol_{#1}}

\newcommand{\MediaMi}[1]{\SourceMediaSymbol{}_{#1}}
\newcommand{\cuticlip}[2]{#1^{#2}_m}
\newcommand{\cutiin}[2]{#1^{#2}_{in}}
\newcommand{\clipiduration}[2]{#1^{#2}_T}
\newcommand{\Cuti}[2]{#1_{#2}}
\newcommand{\cutiout}[2]{#1^{#2}_{out}}

\newcommand{\XTrack}{\textbf{X}}
\newcommand{\xcutinfosymbol}{\text{x}}

\newcommand{\xCuti}[1]{X_{#1}}
\newcommand{\xcutiout}[1]{\xcutinfosymbol{}^{#1}_{out}}

\newcommand{\vcutinfosymbol}{\text{v}}
\newcommand{\vcuticlip}[1]{\vcutinfosymbol{}^{#1}_m}
\newcommand{\vcutiin}[1]{\vcutinfosymbol{}^{#1}_{in}}

\newcommand{\vCuti}[1]{V_{#1}}
\newcommand{\vcutiout}[1]{\vcutinfosymbol{}^{#1}_{out}}

\newcommand{\acutinfosymbol}{\text{a}}
\newcommand{\acuticlip}[1]{\acutinfosymbol{}^{#1}_m}
\newcommand{\acutiin}[1]{\acutinfosymbol{}^{#1}_{in}}

\newcommand{\aCuti}[1]{A_{#1}}
\newcommand{\acutiout}[1]{\acutinfosymbol{}^{#1}_{out}}

\newcommand{\timei}[1]{t_{i}}

\newcommand{\AudioTrackSymbol}{A}
\newcommand{\AudioTrack}{\mathbf{\AudioTrackSymbol{}}}
\newcommand{\AudioCuti}[1]{\Cuti{\AudioTrackSymbol}{#1}}

\newcommand{\VideoTrackSymbol}{V}
\newcommand{\VideoTrack}{\mathbf{\VideoTrackSymbol{}}}

\newcommand{\VideoCuti}[1]{\Cuti{\VideoTrackSymbol}{#1}}

\newcommand{\constrainttype}[1]{\item\textit{#1}}

\newcommand{\ContinuityC}{Continuity}
\newcommand{\SynchronyC}{Synchrony}

\newcommand{\audiovisualsync}{audio-visual synchrony}
\newcommand{\audiovisual}{audio-visual}

\newcommand{\newterm}[1]{\emph{#1}}

\newcommand{\microedit}{micro-edit}
\newcommand{\microediting}{\microedit{}ing}
\newcommand{\microedits}{\microedit{}s}
\newcommand{\Microedit}{Micro-edit}

\newcommand{\MicroEdit}{Micro-Edit}
\newcommand{\MicroEditing}{\MicroEdit{}ing}
\newcommand{\Microedits}{\Microedit{}s}
\newcommand{\MicroEdits}{\MicroEdit{}s}

\newcommand{\macroedit}{macro-edit}
\newcommand{\macroediting}{\macroedit{}ing}
\newcommand{\macroedits}{\macroedit{}s}
\newcommand{\Macroedit}{Macro-edit}
\newcommand{\MacroEdit}{Macro-Edit}
\newcommand{\Macroediting}{\Macroedit{}ing}
\newcommand{\MacroEditing}{\MacroEdit{}ing}
\newcommand{\Macroedits}{\Macroedit{}s}

\newcommand{\CutTypeItem}[1]{\item\textbf{#1}}

\newcommand{\cuttypeend}{-cut}
\newcommand{\LCut}{L\cuttypeend{}}
\newcommand{\JCut}{J\cuttypeend{}}

\newcommand{\Straightcut}{Straight cut}

\newcommand{\splitcut}{split cut}

\newcommand{\Rhythm}{Rhythm}
\newcommand{\rhythm}{rhythm}

\newcommand{\View}{View}
\newcommand{\view}{view}

\newcommand{\EditWindow}{W}
\newcommand{\EditWindowR}{\EditWindow{}_{\text{r}}}
\newcommand{\EditWindowV}{\EditWindow{}_{\text{v}}}

\newcounter{formativefindingcounter}
\newcounter{designgoalcounter}
\iflatexml
\newcommand{\formativefinding}[2]{%
\stepcounter{formativefindingcounter}%
\expandafter\xdef\csname microedits@#2\endcsname{\theformativefindingcounter}%
\namedparagraph{\hypertarget{#2}{F\theformativefindingcounter}) #1}}
\newcommand{\FormFind}[1]{\hyperlink{#1}{\textbf{F\csname microedits@#1\endcsname}}}
\newcommand{\designgoal}[2]{%
\stepcounter{designgoalcounter}%
\expandafter\xdef\csname microedits@#2\endcsname{\thedesigngoalcounter}%
\namedparagraph{\hypertarget{#2}{DG\thedesigngoalcounter}) #1}}
\newcommand{\DesignGoal}[1]{\hyperlink{#1}{\textbf{DG\csname microedits@#1\endcsname}}}
\else
\newcommand{\formativefinding}[2]{\namedparagraph{\refstepcounter{formativefindingcounter}F\theformativefindingcounter\label{#2}) #1}}
\newcommand{\FormFind}[1]{\textbf{F\ref{#1}}}
\newcommand{\designgoal}[2]{\namedparagraph{\refstepcounter{designgoalcounter}DG\thedesigngoalcounter\label{#2}) #1}}
\newcommand{\DesignGoal}[1]{\textbf{DG\ref{#1}}}
\fi

\begin{document}

\newcommand{\PaperTitle}{Micro-Edits: Computational Support for  Fine-Scale Edits in Film}
\newcommand{\ShortPaperTitle}{\PaperTitle{}}

\title[\ShortPaperTitle{}]{\PaperTitle{}}

\author{Nhan Tran}
\iflatexml\else
\authornote{This work was done during the author's internship at Adobe Research.}
\fi
\affiliation{%
  \institution{Cornell University}
  \city{Ithaca}
  \state{New York}
  \country{USA}}
\email{nhan@cs.cornell.edu}

\author{Hijung Valentina Shin}
\affiliation{%
  \institution{Adobe Research}
  \city{Cambridge}
  \state{Massachusetts}
  \country{USA}}
\email{vshin@adobe.com}

\author{Abe Davis}
\affiliation{%
  \institution{Cornell University}
  \city{Ithaca}
  \state{New York}
  \country{USA}}
\email{abedavis@cornell.edu}

\author{Mackenzie Leake}
\affiliation{%
  \institution{Adobe Research}
  \city{San Francisco}
  \state{California}
  \country{USA}}
\email{leake@adobe.com}

\begin{abstract}

In film editing, every frame matters. While macro, content-level edits shape the high-level narrative of a film, small local edits often control timing and direct visual attention. For example, editors can adjust the precise timing of clips to control the length of a meaningful silence, or cut to shots that highlight a specific character's reactions to important moments. These small, detailed edits are slow and tedious to make in standard editing tools because users have to map high-level narrative goals to low-level frame-by-frame adjustments. At the same time, these fine-scale adjustments, which we call \defineterm{\microedits{}}, can be nuanced and expressive, with different choices serving different creative goals. In this work, we start by conducting a formative study with professional editors to understand the structure of \microedits{} and how they are used. We then derive a mathematical formalization for this structure to develop tools that leverage AI and optimization to make exploring \microedits{} simpler and more efficient for users. Finally, we validate our tools in multiple user studies and a wide range of video examples.

\end{abstract}



\begin{teaserfigure}
\includegraphics[width=\textwidth]{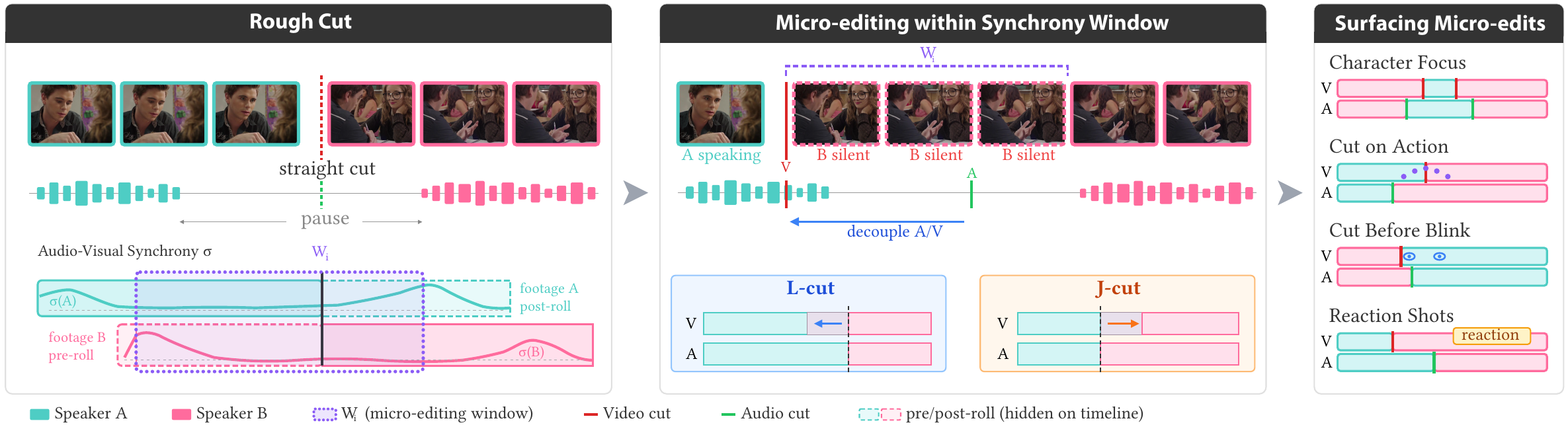}
\caption{After a rough cut via macro-level clip selections yielding straight cuts with coupled audio-video (left), editors refine cuts through \emph{\microedits{}} (center): \rhythm{} edits adjust timing (e.g., pauses), while \view{} edits redirect visual attention. We formalize \emph{\microedits{}} as continuous optimization within \audiovisual{} synchrony-constrained windows, where audio-video can be safely decoupled. Our interactive tools leverage optimization and AI to support different creative objectives (right): automatically balancing character focus, and surfacing perceptual cues for cut-on-action, cut-before-blink, and reaction shots. All video frames shown in this paper are copyrighted and belong to EditStock.\\Project page: {\color{ACMBlue}\url{https://microedits.github.io/}}}
\Description{A three-panel diagram illustrating the transition from a rough cut to specific AI-assisted micro-edits. The left panel shows a straight cut where Speaker A's teal video and audio tracks transition simultaneously to Speaker B's pink tracks. A purple dotted box denotes the micro-editing window over a pause in dialogue. The center panel shows the video cut shifted left within this window, decoupling it from the audio cut. Below, inset diagrams visually contrast an L-cut (video track transitions before audio) and a J-cut (video track transitions after audio). The right panel shows timeline visualizations for four micro-edits: Character Focus, Cut on Action (aligning dot markers), Cut Before Blink (using eye icons), and Reaction Shots (showing an overlapping yellow video segment).}
\label{fig:teaser}
\end{teaserfigure}
\maketitle

\iflatexml
\noindent\textit{Author note (Nhan Tran): This work was done during the author's internship at Adobe Research.}\par
\fi

\section{Introduction}
\label{sec:intro}
Video editing happens at multiple temporal scales, and editors tend to work in a coarse-to-fine manner. At the coarse scale, they begin with what we call \defineterm{\macroedits{}}, which determine the selection and ordering of video clips. \Macroedits{} establish a basic timeline to create what is commonly referred to as a \emph{rough cut} \cite{o2009invisible}. While novice filmmakers and AI-based editing tools often limit editing to rough cuts, skilled filmmakers typically make additional fine-scale edits by decoupling audio and visual transitions and shifting them along a given timeline to control pacing and direct audience attention. These fine-scale edits, which we call \defineterm{\microedits}, offer significant expressive control over a scene's narrative\nchange{~\cite{chandler2012cut, murch2025suddenly}}.
However, they require expertise to execute well and remain tedious even for skilled editors\nchange{~\cite{dmytryk2018film}}.
In this work, we address these challenges by developing new interactive tools to make \microedits{} easier to explore and apply\nchange{, focusing on dialogue scenes, which are especially common in narrative film~\cite{cutting2021movies}}. Our work comprises \nchange{four} parts:

\begin{packed_item}
\item First, we examine how and why expert filmmakers use \microedits{} through a formative user study, detailed in \Sec{\ref{sec:formative_concise}}. We found that experts use \microedits{} in two distinct ways: to manipulate the timing of a narrative (\rhythm{} edits), and to direct visual attention (\view{} edits).
\item Next, we formalize the space of possible \microedits{} for a given scene in \Sec{\ref{sec:microstructure}}, and use our formulation to describe different editing constraints and objectives. This allows us to address \microediting{} as an optimization problem and incorporate analyses provided by vision-language models (VLMs), when available.   
\item \nchange{Then}, building on our formalization, we design novel interactive tools that leverage optimization and AI to efficiently explore \microedits{} through interactions that map more clearly and consistently to common creative goals.
\item \nchange{Finally, we validate our tools in \Sec{\ref{sec:evaluation}} through a crowdsourced study and two evaluations with 16 experienced editors, showing that they reduce mechanical effort and support broader exploration of \microedits{}.}
\end{packed_item}

Our work combines user-centered design with the development of novel technical tools. The design contributions in our work deal with understanding how \microedits{} are used so we can build better editing tools, while our technical contributions address challenges that emerge when applying computation to that understanding. \nchange{Our formative findings translate into two design goals: efficient exploration and useful suggestions (\Sec{\ref{sec:design_goals}}). We use these goals to motivate and evaluate our tools.}

\section{Related Work}
\label{sec:relatedwork}

While our work is, to our knowledge, the first to address interactive tools for \microediting{}, prior work in vision, graphics, and HCI has looked at related problems, including rough cut assembly.

\subsection{Macro-Level Editing and Assembly}
Computational tools for video editing primarily target content selection and rough-cut assembly. Systems for scripted dialogue encode cinematographic idioms to decide which clips appear and in what order \cite{leake2017computational, leake2024chunkyedit}. Transcript-based interfaces \cite{truong2016quickcut, casares2002simplifying, pavel2015sceneskim} enable text-driven manipulation, while keyword analysis recommends B-roll insertion points \cite{huber2019b}. More recent LLM-powered systems infer editorial intent from natural language \cite{wang2024lave}, combine language with sketch-based control \cite{tilekbay2024expressedit}, and support human–AI co-creation for generating rough-cuts \cite{sandoval2025editduet, huh2025videodiff}. \nchange{Recent work on automated montage assembly learns shot selection, trimming, and cut placement from professionally edited films, but does not provide explicit editor control over \microedits{}, including decoupled audio and video transitions~\cite{sandoval2026filmgpt}.} These systems operate at word- or shot-level granularity, which is effective for structuring a scene, but they stop short of the frame-level refinement where professional editors shape timing and attention.

\subsection{Automated Cut Placement and Pacing}

Commercial tools provide filler-word removal and silence compression to streamline rough cuts~\cite{descript,opusclip2025,capcut2025}. Adobe Premiere plugins like AutoCut~\citeyearpar{autocut2025} and FireCut~\citeyearpar{firecut2025} extend this by offering configurable offsets within silence-removal workflows, where audio is cut earlier than video by a fixed amount to smooth transitions (split edits). These workflows treat pacing as uniform temporal compression and treat split edits as mechanical timing offsets. Such behavior is well-suited to high-volume social and podcast workflows with strict duration constraints and tolerance for jump-cut rhythm, but misaligned with dramatic editing where silence is expressive and overlap is applied selectively to anticipate or extend specific beats. Our work formalizes these micro-editing operations and introduces an interactive tool for selective, frame-level refinement of rhythm and visual emphasis after rough-cut assembly. Rather than applying uniform pacing or fixed offsets, our system surfaces alternatives and perceptual cues tied to narrative beats and viewer attention.

The closest prior work at comparable granularity is \citet{berthouzoz2012tools}, who analyze audio-visual features in single-speaker interview footage to locate frames where the subject is quiet and still, enabling hidden transitions via graph-based frame selection. Our work differs in three respects: (1) we address multi-clip dialogue with multiple speakers and camera angles; (2) we support audio–video decoupling through split edits rather than solely audio-aligned straight cuts; and (3) we provide interactive exploration of alternatives rather than optimizing for a single global solution.

\subsection{Micro-Temporal Signals for Editing}
Editors time cuts to perceptual events that guide attention and maintain continuity~\cite{smith2012attentional,smith2013attentional}. Conventions such as cutting mid-action~\cite{dmytryk2018film}, decoupling picture and sound cut points, and avoiding cuts on blinks exploit motion continuity, dialogue structure, and \audiovisualsync{}~\cite{murch2025suddenly,salt2011reaction}. \citet{galvane2015continuity} encode some continuity constraints over screen position, gaze, and motion, and optimize shot sequences accordingly in virtual environments where scene state is available. In practice, these constraints often compete, and their relative importance depends on narrative context.

In real footage, these cues can be inferred from \audiovisual{} signals, including gaze and blink via facial landmark tracking~\cite{lugaresi2019mediapipe} and \audiovisualsync{}~\cite{owens2018audio, sun2023eventfulness}. Prior work aligns or matches video using appearance similarity \cite{wang2014videosnapping}, motion trajectories or optical flow for action synchronization \cite{arev2014automatic, bazin2016actionsnapping, chen2023match}, and learned features for continuity scoring and transition classification \cite{pardo2021learning, pardo2022moviecuts}.

Semantics can further indicate which alignments are editorially salient. Recent VLMs, when supplied with numbered-frame inputs~\cite{wu2025number} and short, densely sampled clips~\cite{qu2025does}, can identify such moments (e.g., a door slam, a character's turn). We build on these semantic and \audiovisual{} cues to surface micro-edit opportunities within an interactive workflow.

\section{Background and Formative Study}
\label{sec:formative_concise}

We conducted a formative study with six experienced editors (5--15 years experience) to understand how and why they adjust rough cuts. Full details of our study are included as supplemental material, and we summarize our findings below.

\subsection{Background}
Our study design incorporates additional background related to how editors often discuss \emph{cuts}, which represent transitions on a timeline (video, audio, or both) from one piece of source media to another.
Note that the term ``cut'' has two meanings in editing: it can refer to a \emph{transition point} between clips, or to a \emph{version} of the edit (as in ``rough cut'' or ``final cut''). Here we discuss cuts as transition points.
Editors classify most cuts into one of three categories:
\begin{packed_item}
\CutTypeItem{\Straightcut{}s} happen when video and audio transition together.
\CutTypeItem{\LCut{}s} happen when video transitions before audio.
\CutTypeItem{\JCut{}s} happen when audio transitions before video.
\end{packed_item}
Additionally, the term ``\splitcut{}'' is sometimes used to describe any cut that is not a straight cut. Implicit in these categories is an assumption that both audio and video transition between corresponding source media. While not always true, this describes most cuts found outside specific genres like music videos. 

\Macroediting{} is often limited to straight cuts, and previous computational tools have made this an explicit assumption when formulating rough cut assembly \cite{leake2017computational, truong2016quickcut, huh2025videodiff}. This points to L- and J-cuts as offering some insight into how editors think about \microediting{}, which we leverage in our formative study design.

\subsection{Study Design}
 Each participant was given rough cuts of two scripted scenes (each approximately one-minute long) and tasked with editing each scene in an NLE of their choice under four conditions. In one condition, participants were unrestricted. For each of the remaining three conditions, we asked participants to limit edits to different cut types: straight cuts, L-cuts, or J-cuts. Participants were recorded and asked to explain their reasoning aloud while editing, and after completing their tasks, each participant filled out a survey.

\subsection{Formative Findings}

\formativefinding{Every Frame Matters}{ff:everyframe} Our study reinforced the common claim that skilled editors often make decisions at the scale of individual frames\nchange{~\cite{murch2001blink, dmytryk2018film}}. While making \microedits{}, participants described ``\textit{going frame by frame, removing one frame here, adding a few there}'' \nchange{(P5)} or ``\textit{adding 3 or 4 more frames}'' \nchange{(P1)} to get transitions to feel just right. This process also involved repeatedly re-watching parts of video to compare the feel of different choices.

\formativefinding{Contextual Information is Important}{ff:context}  
Editing decisions often depend on local context. For example, editors generally place audio cuts in silent parts of a clip. On the other hand, they often adjust visual cuts based on narrative goals (e.g., cutting to see a particular character's reaction during a dramatic moment), or following the ``cut on action'' principle \cite{dmytryk2018film, chandler2012cut}, where visuals before and after a cut should show the continuity of some visible action (e.g., a gesture, blink, or gaze turn). This type of higher-level reasoning helps editors narrow their choice of edit.

\formativefinding{Control Over Rhythm \& Pacing}{ff:rhythm}
Editors often start with edits meant to control the precise rhythm or pacing of a scene---usually by inserting or removing time at the ends of adjacent clips. We call this type of adjustment a \newterm{\rhythm{} edit}. \Rhythm{} edits change how events are distributed on the timeline; for example, in dialogue-heavy scenes, they can be used to control the length of pauses in conversation. \nchange{Half of our participants} described focusing exclusively on the timing of dialogue while making such edits, with one \nchange{(P6)} describing this process as starting with a ``\textit{radio edit}'', and another \nchange{(P4)} saying ``\textit{I'm going to close my eyes and listen ... this lets me really focus on the flow and not be interrupted by some video breaks.}''
Rhythm edits require shifting a clip along the output timeline, often in addition to adjusting the duration of a clip.

\formativefinding{Control Over Visual Attention}{ff:views}
Later edits often focus on controlling visual attention. We can think of these edits, which we call \newterm{\view{} edits}, as controlling the viewpoint of the audience without changing the timeline of events that are happening. \View{} edits are where participants tended to make \LCut{}s and \JCut{}s. More specifically, \LCut{}s were mostly used to make visual transitions happen sooner, while \JCut{}s were used to make them happen later. In both cases, this was usually done without changing the audio track. To quote one participant \nchange{(P1)}, ``\textit{I already changed the timing of the audio... so I don't wanna, like, mess with that.}''

\formativefinding{Inserting New Cuts For Reactions \& Cutaways}{ff:reactions}
In our study, most edits involved adjusting existing cuts. The main exception to this happened when editors wanted to add a reaction or cutaway shot, where visuals transition away from the speaker or current source of audio. Two reasons were given for such edits: first, to direct visual attention; and second, to mask jump cuts (created when trimming pauses within dialogue). Also, editors often repurpose visuals from other parts of a scene to create reaction shots. This points to conditions where reordering clips in a scene is acceptable.

\subsection{Design Insights}
\FormFind{ff:everyframe} and \FormFind{ff:context} reflect that \microediting{} does not have a unique solution, and precise user control is important, though evaluating contextual and semantic information may help automate some decisions.

\noindent\FormFind{ff:rhythm} and \FormFind{ff:views} suggest that editors treat \rhythm{} and \view{}s separately; thus, controls for these aspects should be independent. This separation reflects how editors exploit \emph{\audiovisualsync{}}, as one participant \nchange{(P1)} articulated: ``\textit{What you hear and what you see are two different systems... if you introduce the hearing as one sense and then introduce the sight, it's a different feeling than if you introduce both at the same time.}'' Our tools are designed to help users focus on each type of control separately. \FormFind{ff:reactions} suggests that showing only speakers means audiences miss half of the story---character reactions---so editors spend significant time scrubbing through takes to find them. A tool that surfaces candidates would help.

\subsection{Design Goals}
\label{sec:design_goals}
\nchange{Different \microedits{} may accomplish different creative goals, which can be highly context-dependent. This lack of a single correct solution complicates automation, but there is still a great deal that optimization and AI can offer to improve existing workflows.} \Microedits{} are not random, and the space of all possible edits is much larger than the set that meet likely creative goals. With this in mind, we focus on two overarching design goals in this work:

\designgoal{More Efficient Exploration}{dg:exploration} \nchange{Our findings (\FormFind{ff:everyframe}) illustrate} that much of the challenge in making \microedits{} comes from a ``gulf of execution'' \cite{NormanDesignOfEverydayThings} between changes users want to make and the steps those changes require in existing non-linear editors (NLEs). Our tools should make \microedits{} easier to execute when possible, letting users explore the same edit space more efficiently.

\designgoal{Useful Defaults \& Suggestions}{dg:suggestions} \nchange{As we discuss in \Sec{\ref{sec:microstructure}}, this space is continuous,} but editors often reason about \microedits{} in ways that give rise to more discrete sets of likely editing decisions. Our tool should help identify such decisions and present them to users as defaults or suggestions.

These design goals reflect a need to balance automation with expressive control: \nchange{our findings identify the types of control that editors want, and our tools focus on automating the work that makes this control less efficient. Meeting these goals requires describing this space precisely, which we do in the next section.}

\begin{figure}
  \begin{center}
  \includegraphics[width=\linewidth]{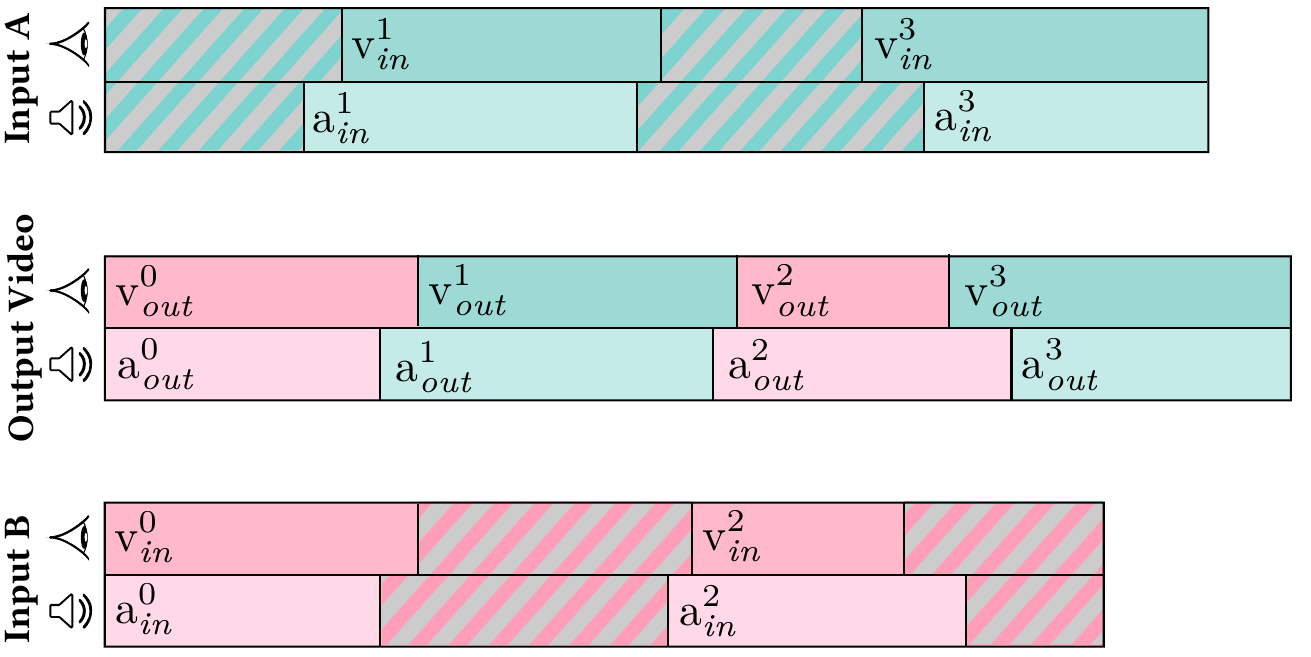}
  \end{center}
    \caption{Edit structure showing how an edited video combines segments from two source media across video and audio tracks. Each cut $\xCuti{i}$ is defined by its output time ($\cutiout{x}{i}$), source media ($\cuticlip{x}{i}$), and source time ($\cutiin{x}{i}$). Striped regions indicate unused footage. Colors indicate source origin.}
  \Description{Edit structure showing how an edited video combines segments from two source media across video and audio tracks. Three stacked timelines display Input A (teal, top), Output Video (middle), and Input B (pink, bottom). Each features a video track marked by an eye icon and an audio track marked by a speaker icon. Input A and B show the original source media, with solid colored blocks for used segments and gray striped blocks indicating unused footage. Input A provides segments 1 and 3, while Input B provides segments 0 and 2. The Output Video timeline shows these segments spliced together sequentially (0, then 1, 2, and 3), alternating between pink and teal. The vertical cut lines between segments in the Output Video are staggered across the video and audio tracks, visually demonstrating that the video and audio transitions occur at slightly different times.}
  \label{fig:editstructure}
\end{figure}

\begin{figure*}
\begin{center}
    \includegraphics[width=0.92\textwidth]{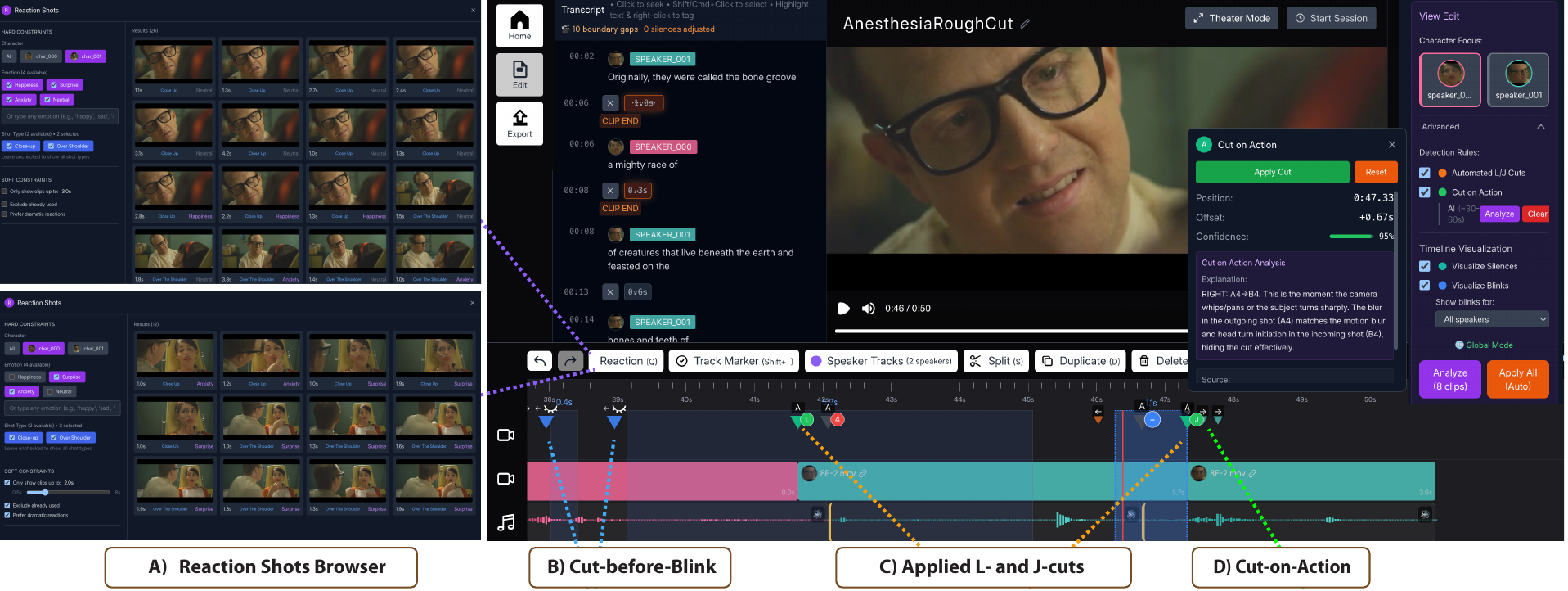}
\end{center}
\caption{\nchange{Our web-based editor provides standard NLE controls alongside specialized \microediting{} tools. Track colors correspond to characters, and blue overlays mark silence regions for \rhythm{} edits, which editors adjust with pre-roll and trim\slash extend hotkeys during playback. For \view{} edits, the Character Focus panel (right) automates L/J-cut placement to emphasize a selected character's screen time, and applied offsets appear on the timeline (C). Markers at each cut surface perceptual-cue suggestions: cut-before-blink candidates (B) and cut-on-action candidates (D). Hovering over a marker reveals an AI-generated explanation; clicking applies or resets the suggested offset. The Reaction Shot Browser (A) retrieves non-speaking moments, filterable by character, emotion, and shot type; editors insert reactions by selecting transcript text or a timeline position.} Video frames~\textcopyright~EditStock.}

    \Description{A screenshot of the web-based video editor interface showing specialized micro-editing tools and timeline. The image is split into a left section displaying two stacked Reaction Shot Browsers (Label A), each showing a grid of video thumbnails next to a side panel of filtering options. The right section shows the main editor, containing a text transcript, a video player, a right-side settings panel, and a multi-track timeline at the bottom. Dotted lines highlight three timeline features: Label B, "Cut-before-Blink," points to a small eye icon marker at a cut; Label C, "Applied L- and J-cuts," points to visually staggered transitions between the colored video and audio tracks; and Label D, "Cut-on-Action," points to a green marker that opens a pop-up dialog box containing an AI analysis explanation and an "Apply Cut" button.}
\label{fig:view_pass_ui}
\end{figure*}

\section{Problem Formulation}
\label{sec:microstructure}

We describe editing as a two-stage process, starting with \macroedits{}, followed by \microedits{}. \nchange{We use this framing to distinguish two types of editing decisions, not to claim that editing proceeds strictly in this order in practice.} The input to this process is a set of source media and optional narrative description. Source media is given as a set $\SourceMediaSet{}=\{\MediaMi{\varnothing},\MediaMi{1},...\}$ of clips, where $\MediaMi{\varnothing}$ represents the empty clip (i.e., what is displayed if part of a timeline is left empty). 
The narrative, if provided (e.g., in the form of a script or storyboard), is represented with a sequence of narrative \element{}s $\NarrativeSet{}=[\NElementi{1},\NElementi{2},...]$, where each $\NElementi{i}$ describes a different part of the narrative. We use $\NElementi{i}\eventin\MediaMi{j}$ to indicate that $\NElementi{i}$ is portrayed in source media $\MediaMi{j}$, and $\MediaMi{j}(\NElementi{i})$ to denote the part of $\MediaMi{j}$ that depicts $\NElementi{i}$. We leave the definition of \narrative{} \element{}s $[\NElementi{1},\NElementi{2},...]$ intentionally loose, as their primary role is to segment source media and express partial orderings during \macroediting{}. We will introduce more precise notation later for discussing narrative events in \microedits{}.

\subsection{Rough Cuts \& \MacroEditing{}} 
\label{sec:representingroughcuts}
The goal of \macroediting{} is to produce a rough cut $\RoughCut{}$, given as an ordered sequence of source clips. When a \narrative{} is provided, $\RoughCut{}$ is determined by a mapping $\roughmap{}$ from each $\NElementi{i}\in\NarrativeSet{}$ to a corresponding clip $\RoughCutClipi{i}$ taken from source media $\MediaMi{\roughmap{}(i)}$:
\beq
\RoughCut=[\RoughCutClipi{1},\;\RoughCutClipi{2},\; ...\;, \;\RoughCutClipi{\nEvents}]
\label{eq:roughcut}
\eeq
What makes this ``rough'' is twofold: clip boundaries are left inexact, and cuts are restricted to straight cuts. From the editor's perspective, \microediting{} refines the former and relaxes the latter.
Mathematically, a rough cut makes the set of potential clips discrete and finite. 
In this case, as $\roughmap{}$ maps from one discrete finite set to another, the space of possible rough cuts becomes discrete and finite. We can represent each $\RoughCut$ in this space as a path through the directed graph of possible clips, where edges encode the ordering constraints given by narrative $\NarrativeSet{}$. If $\NarrativeSet{}$ is not provided, this graph is fully connected, and we need to limit the maximum number of clips to keep the space of rough cuts finite.

\subsection{Representing \MicroEdits{}}
\label{sec:microeditingnotation}
 In a sense, the purpose of \microedits{} is to address assumptions from rough-cut assembly. The input to our \microediting{} tools is a rough cut $\RoughCut{}$ (\EqRef{eq:roughcut}), whether assembled manually or by automatic methods~\cite{leake2017computational, sandoval2025editduet}. From this, \microediting{} refines imprecise clip boundaries and relaxes audio-video coupling.
To do this, we need more precise notation.
Our output will be a pair $\{\VideoTrack{},\AudioTrack{}\}$ of edited \emph{tracks} representing output video and audio, respectively (Figure~\ref{fig:editstructure}):
\baeq
    \VideoTrack{}&=[\vCuti{1}, \dots, \vCuti{\nclips}], \;\;\text{where}\;\; \vCuti{i}=\{\vcutiout{i}, \vcuticlip{i}, \vcutiin{i}\}\\
    \AudioTrack{}&=[\aCuti{1}, \dots, \aCuti{\nclips}], \;\;\text{where}\;\;
    \aCuti{i}=\{\acutiout{i}, \acuticlip{i}, \acutiin{i}\}
\label{eq:trackrepresentation}
\eaeq
\noindent{}More generally, each track $\XTrack{}$ (where $\mathbf{X} \in \{\mathbf{V}, \mathbf{A}\}$) is a sequence of \emph{cuts},
where cut $\xCuti{i}$ denotes a transition between discontiguous parts of source media. Each cut is described as a triplet 
$\Cuti{X}{i}=\{\cutiout{x}{i}, \cuticlip{x}{i}, \cutiin{x}{i}\}$,
where $\cutiout{x}{i}$ denotes when the cut occurs in the output timeline, $\cuticlip{x}{i}\in{}\SourceMediaSet{}$ identifies the \sourcemedia{} being cut to, and $\cutiin{x}{i}$ gives the corresponding source time.
For convenience, we use $\clipiduration{x}{i} = (\cutiout{x}{i+1} - \cutiout{x}{i})$ as shorthand for the duration of $\cuticlip{x}{i}$ on track $\mathbf{X}$.

\subsubsection{Initial Rough Cuts}  In initial rough cuts, audio and video cuts are coupled, meaning $\VideoCuti{i}=\AudioCuti{i}$ for all cuts. The source media for each cut is defined by $\vcuticlip{i}=\acuticlip{i}= \MediaMi{\roughmap{}(i)}$. The output time $\vcutiout{i}=\acutiout{i}$ is set to where clip $\RoughCutClipi{i}$ begins on the output timeline, and input time $\cutiin{x}{i}$ to where it begins in source media $\MediaMi{\roughmap{}(i)}$.

\subsubsection{The Space of \MicroEdits{}}
Starting from a rough cut ensures that most \microedits{} are limited to adjustments of the continuous $\cutiout{x}{i}$ and $\cutiin{x}{i}$ parameters of existing cuts, leaving the ordered sequence of source clips $\cuticlip{x}{i}$ unchanged. We will start by focusing on such adjustments, and consider the addition of new cuts such as reaction shots later. 
As $\cutiout{x}{i}$ and $\cutiin{x}{i}$ represent continuous time offsets, the space of possible \microedits{} is continuous.

\subsection{\MicroEditing{} Constraints \& Context}

We identify two types of constraints that impact \microediting{}: 
\begin{packed_item}
    \constrainttype{\SynchronyC{} constraints} ensure that visible sound sources are heard (e.g., we hear a character who is visibly speaking). This constraint is one-directional: hearing off-screen sound is acceptable, which enables L- and J-cuts. We quantify this via \synchrony{}~\cite{desynchronydetection,roseboom2009sliding,vatakis2006audiovisual}: audio and video should stay coupled in high-synchrony regions. 
    \constrainttype{\ContinuityC{} constraints}  maintain consistency across cuts. For example, when cutting to different viewpoints of the same scene, the poses and appearance of characters should remain largely unchanged. Similarly, when cutting to different audio of the same scene, one should avoid cutting in the middle of salient sound, as this can create audible discontinuities.  
\end{packed_item}
Both of these types deal with some types of consistency: between audio and video, or across cuts. Strictly speaking, there are no editing patterns that these constraints prevent in all scenarios; each depends on the context of a particular scene. To help reason about this, we need a way to represent context in source media. Unlike the coarse narrative elements $\NarrativeSet{}$ used for \macroediting{}, \microediting{} requires fine-grained events---blinks, action onsets, reactions, word boundaries. We denote these $\EventSet{}=[\eventi{1},\eventi{2},...]$, with $\MediaMi{j}(\eventi{i})$ denoting the time in $\MediaMi{j}$ where $\eventi{i}$ occurs. The key context function is synchrony $\synchronyfunc{}(\MediaMi{j}, t)$, representing audio-visual alignment at time $t$. The next section describes how our system uses these constraints and context to support \microediting{} tools.

\section{System Design}
\label{sec:system_design}

\begin{figure}[t]
\centering
\includegraphics[width=1.0
\columnwidth]{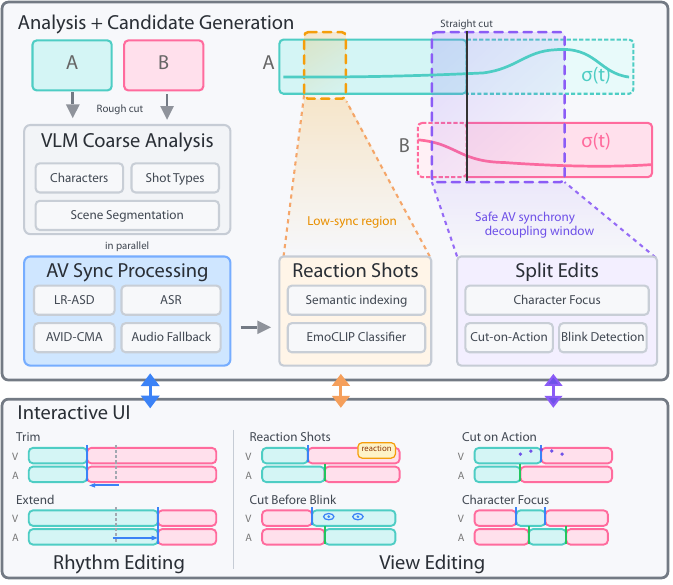}
\caption{
System overview: From a rough cut, our backend performs analysis
(VLM, AV synchrony, reaction-shot indexing, cut-on-action, blinks) and surface cut suggestions on UI.
}
\Description{System overview flowchart showing how a rough cut is analyzed to generate cut suggestions for the UI. The diagram is divided into a top "Analysis + Candidate Generation" backend and a bottom "Interactive UI" frontend. In the backend, source clips A and B are processed in parallel by VLM Coarse Analysis and AV Sync Processing. These analyses identify timeline constraints: a "Low-sync region" that generates "Reaction Shots" (using semantic indexing and EmoCLIP), and a "Safe AV synchrony decoupling window" that generates "Split Edits" (including Character Focus, Cut-on-Action, and Blink Detection). Bi-directional arrows connect these generated candidates to the Interactive UI below. The UI section displays the resulting tools, split into "Rhythm Editing" (showing timeline visualizations for Trim and Extend operations) and "View Editing" (showing visualizations for Reaction Shots, Cut Before Blink, Cut on Action, and Character Focus).}
\label{fig:system_overview}
\end{figure}

Our system provides tools for \rhythm{} edits (\Sec{\ref{sec:rhythm}}), \view{} edits (\Sec{\ref{sec:view}}), and reaction shots (\Sec{\ref{sec:reactionshots}}).
Figure~\ref{fig:system_overview} presents our end-to-end architecture.
Given source media $\SourceMediaSet{}$ and a rough cut $\RoughCut{}$, the backend runs analysis and candidate generation in parallel, and the Interactive UI supports human-in-the-loop refinement via rhythm and view edits.
Following \Sec{\ref{sec:microeditingnotation}}, \rhythm{} and \view{} edits operate on continuous parameters $(\cutiout{x}{i}, \cutiin{x}{i})$ while leaving $\cuticlip{x}{i}$ unchanged, and reaction shots insert additional clips on an overlay track.
\nchange{Our system separates analysis from interaction: an offline pipeline annotates source clips, while the editing interactions are implemented directly from our formalization (\Sec{\ref{sec:implementation_details}}).}

\subsection{Editing Windows}
\label{sec:syncwindows}
At each cut $\xCuti{i}$, we transition from an \emph{outgoing} clip $\cuticlip{x}{i-1}$ to an \emph{incoming} clip $\cuticlip{x}{i}$. Each clip's source footage usually extends beyond what is currently visible: the outgoing clip has unused \emph{post-roll}, and the incoming clip has unused \emph{pre-roll}. Adjusting a cut reveals footage from these regions. We define an \emph{editing window} $\EditWindow{} = [-\Delta^{-},\; +\Delta^{+}]$ (Figure~\ref{fig:teaser}, bottom left), where negative values reveal incoming pre-roll and positive values reveal outgoing post-roll:

\begin{packed_item}
\item \textbf{\Rhythm{} edits} use $\EditWindowR{}$: adjustments refine pacing while keeping spoken content unchanged. $\Delta^{-}$ extends until the incoming clip's pre-roll would reveal additional speech; $\Delta^{+}$ extends until the outgoing clip's post-roll would reveal additional speech.
\item \textbf{\View{} edits} use $\EditWindowV{}$: $\Delta^{+}$ extends to the first frame where $\synchronyfunc{}(\cuticlip{x}{i-1}, t) = 1$ in post-roll; $\Delta^{-}$ extends to where $\synchronyfunc{}(\cuticlip{x}{i}, t) = 1$ in pre-roll.
\end{packed_item}

\noindent For dialogue scenes, these constraints often align: speakers are silent between turns, and silent speakers have low synchrony. However, the distinction matters when lips continue moving after the last word, or when non-speech sounds create synchrony without words.

\subsection{\Rhythm{} Edits}
\label{sec:rhythm}                                                    

\Rhythm{} edits adjust cut timing while maintaining track coupling ($\VideoCuti{i} = \AudioCuti{i}$). Repositioning a cut changes both $\cutiout{x}{i}$ and $\cutiin{x}{i}$, pulling dialogue earlier (time-contracting) or pushing it later (time-expanding) without cutting into speech. As noted in \FormFind{ff:rhythm}, editors often focus on temporal flow when adjusting rhythm, yet existing NLEs require spatial manipulation of timeline elements. To address this, we provide two hotkey interactions: (1) \emph{pre-roll} hotkey rewinds before the current cut to provide conversational context. (2) A \emph{trim\slash extend} hotkey repositions the cut so that the first word of the incoming clip (following $\xCuti{i+1}$) lands at the current playhead position $p$. If $p < \xcutiout{i+1}$, the system advances the cut (time-contracting, closing the silence gap). 
If $p > \xcutiout{i+1}$, the system delays the cut (time-expanding, revealing silence).
This enables \emph{marking timing by feel}: \citet{murch2001blink} describes marking a shot's cut point in real time and treating his ability to consistently ``hit the same frame'' as confirmation that the timing is right, adjusting until he can~\cite{murch2025suddenly}.

\subsection{\View{} Edits}
\label{sec:view}

\View{} edits decouple video from audio, shifting the video cut by offset $\delta_i$ relative to audio: $\vcutiout{i} = \acutiout{i} + \delta_i$ and $\vcutiin{i} = \acutiin{i} + \delta_i$. An offset $\delta_i < 0$ produces an L-cut; $\delta_i > 0$ produces a J-cut. The constraint $\delta_i \in \EditWindowV{}$ ensures offsets stay in low-synchrony regions.

\begin{figure}[h]
\begin{center}
    \includegraphics[width=0.9\columnwidth]{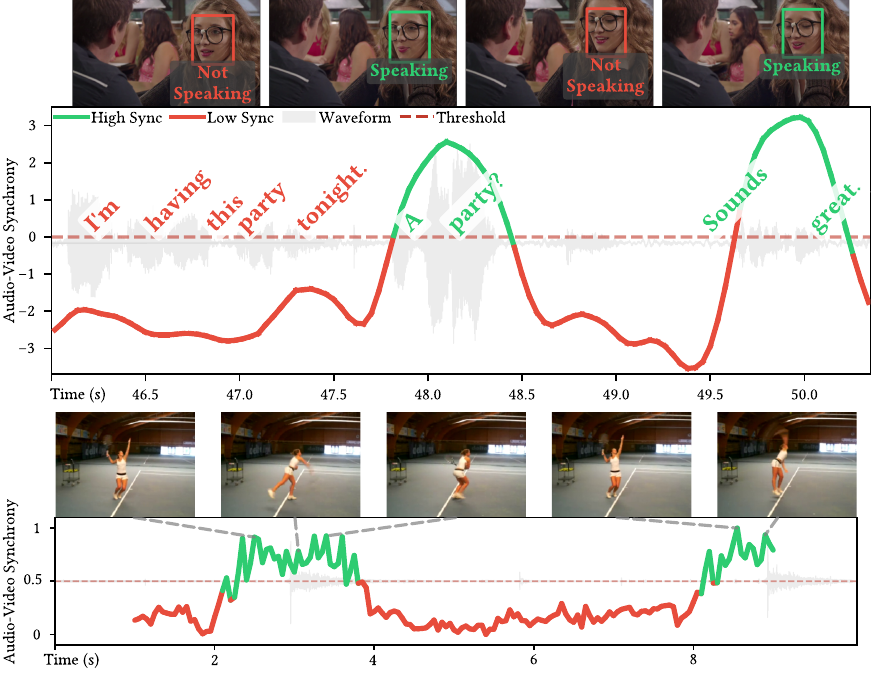}
\end{center}
    \caption{Proxy for audio-visual synchrony ($\synchronyfunc{}$). Top: active speaker detection for dialogue scenes---high synchrony (green) when speaking, low (red) when silent/off-screen. Bottom: general synchrony network for non-dialogue content (tennis). Both guide safe decoupling for L/J-cuts and reaction shots. Video frames~\textcopyright~\mbox{EditStock}.}
    \Description{Two line graphs demonstrating proxy models for audio-visual synchrony in dialogue and non-dialogue scenes. The top graph shows active speaker detection. Four video frames of a woman correspond to a synchrony curve plotted against time, with a background audio waveform. A horizontal threshold line at zero separates low synchrony (red line, matching "Not Speaking" frames) from high synchrony (green peaks, matching "Speaking" frames). Spoken words track along the curve: it remains red below the threshold during the phrase "I'm having this party tonight," spikes green above the threshold for "A party?", dips red again, and spikes green for "Sounds great." The bottom graph analyzes non-dialogue content, featuring a tennis player. Five video frames map to a similar curve with a 0.5 threshold. The curve spikes green above the threshold twice, corresponding exactly to the moments the player's racket strikes the ball, and remains red during the movement between hits.}
    \label{fig:av_synchrony}
\end{figure}

Rather than requiring editors to manually search for optimal offsets within $\EditWindowV{}$, we surface favorable candidates according to established editing patterns in three modes:

\namedparagraph{Character Focus} This optimizes $\delta_i$ across cuts to emphasize a selected character's screen time. Let $d_i \in \{-1, +1\}$ encode whether to favor the outgoing ($d_i = +1$) or incoming ($d_i = -1$) speaker at $\xCuti{i}$. 
\[
\delta_i^* = \arg\min_{\delta \in \EditWindowV{}}
\Bigl(-d_i\,\delta
+ \lambda_{\text{sync}}\,C_{\text{sync}}(\delta)
+ \lambda_{\text{dur}}\,C_{\text{dur}}(\delta)
\Bigr),\]
where $C_{\text{sync}}(\delta) = \synchronyfunc{}(\MediaMi{j}, t_{\text{cut}})$ (the synchrony at the video cut frame), and $C_{\text{dur}}(\delta)$ penalizes offsets that would trim an adjacent clip too short.
We set $\lambda_{\text{sync}} = 0.7$, biasing offsets away from high-synchrony frames at the cut, and $\lambda_{\text{dur}} = 100$, heavily penalizing transitions that would yield flash frames or uncomfortably short clips.
This provides a starting point that editors can refine (\DesignGoal{dg:suggestions}).

\begin{figure}
  \centering
  \includegraphics[width=\linewidth]{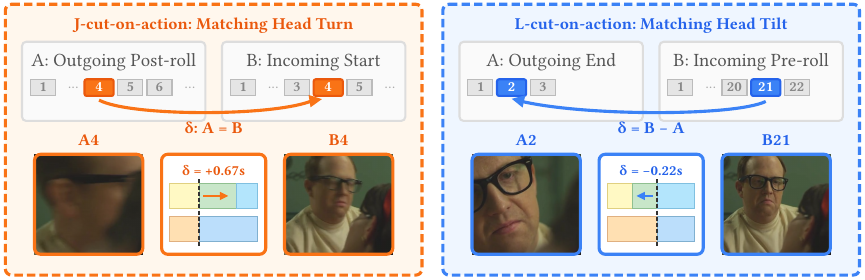}
\caption{Numbered frame mosaics for VLM-based cut-on-action alignment. \textbf{J-cut} ($\vcutiout{i} > \acutiout{i}$): A (unused post-roll) and B (visible start) extend \emph{forward} from the straight cut. Equal indices = same moment, so A4$\to$B4 $\Rightarrow$ $+0.67$s. \textbf{L-cut} ($\vcutiout{i} < \acutiout{i}$): A (visible end) and B (unused pre-roll) extend \emph{backward} from the straight cut. A has 3 frames (short clip), B has 22 (longer footage), so matching aligns by counting backward: B21$\to$A2$\Rightarrow$ $-0.22$s. Video frames~\textcopyright~EditStock.}
  \Description{Diagram illustrating how matching frame numbers between two video clips calculates the timing offset for L-cuts and J-cuts. The left side explains a J-cut for a matching head turn by comparing frames extending forward from the original straight cut point. Since both clips start counting from the same moment, matching frame 4 of the outgoing clip (Clip A) to frame 4 of the incoming clip (Clip B) aligns the action, resulting in a positive 0.67-second offset. The right side explains an L-cut for a matching head tilt, comparing frames extending backward from the cut point. Because the clips have different lengths (Clip A has 3 frames, Clip B has 22), the frame numbers differ. Aligning frame 21 of the incoming clip back to frame 2 of the outgoing clip aligns the action by counting backward, creating a negative 0.22-second offset.}
  \label{fig:cut_on_action_mosaic}
\end{figure}

\namedparagraph{Cut-on-Action}
This mode surfaces candidates where motion continues smoothly across the cut (Figure~\ref{fig:view_pass_ui}D). Since VLMs struggle with micro-temporal reasoning~\cite{qu2025does}, we densely sample frames from the handles (outgoing post-roll and incoming pre-roll) and assemble them into numbered mosaics (Figure~\ref{fig:cut_on_action_mosaic}). Explicit frame indices improve temporal grounding~\cite{wu2025number}. 
The VLM identifies frame pairs with salient motion continuity (e.g., matching gesture trajectory, gaze direction), and indices are converted to offsets $\delta_i$ within $\EditWindowV{}$. \nchange{Restricting candidates to $\EditWindowV{}$ means the VLM reasons over a few synchrony-safe frames rather than the full footage.}

\namedparagraph{Blink Detection}\mbox{}\\[-0.6\baselineskip]

\noindent
\begin{minipage}[t]{0.75\columnwidth}
\raggedright
MediaPipe~\cite{lugaresi2019mediapipe} estimates eye aspect ratio (EAR) per frame from face mesh landmarks. Blink onsets occur when EAR falls below threshold for consecutive frames. As shown in Figure~\ref{fig:view_pass_ui}B, we surface candidates $\delta_i$ that place the video cut just before detected blink onsets, constrained to $\EditWindowV{}$.
\end{minipage}\hspace{0.02\columnwidth}%
\begin{minipage}[t]{0.20\columnwidth}
\vspace{-0.6\baselineskip} %
\raggedleft
\includegraphics[width=\linewidth]{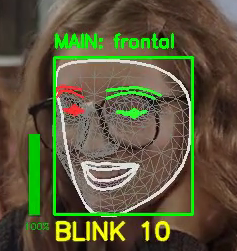}
\end{minipage}

\subsection{Reaction Shots}            
\label{sec:reactionshots}

Unlike \rhythm{} and \view{} edits, which adjust existing cuts, reaction shots add new visual material on an overlay track (B-roll) without altering the primary audio timeline (\FormFind{ff:reactions}). We identify candidate segments where $\synchronyfunc{}(\MediaMi{j}, t) = 0$ (subject visible but not speaking), the target subject is detected in at least 80\% of frames, and the segment is adjacent to spoken material. To support browsing, candidates are indexed by subject identity and emotion using EmoCLIP~\cite{foteinopoulou2024emoclip}. The interface groups results by character and emotion (Figure~\ref{fig:view_pass_ui}A). This supports both expressive goals (showing a character's reaction) and practical needs (masking jump cuts from \rhythm{} tightening).

\subsection{Implementation}
\label{sec:implementation_details}
\nchange{We implemented \emph{MicroEditor} as a web-based editor forked from ClipJS~\cite{clipjs2024}, extending it with multi-track timeline support, rolling edits, and the tools described above.}
An offline pipeline preprocesses each source clip to extract word-level timestamps (via ASR~\cite{radford2023robust}), per-frame synchrony scores, and subject identity. For synchrony, we use active speaker detection~\cite{liao2025lr} for dialogue scenes and an audio-visual alignment network~\cite{morgado2021audio} for non-dialogue content. We threshold the continuous signal to obtain $\synchronyfunc{}(\MediaMi{j}, t) \in \{0, 1\}$, where $\synchronyfunc{} = 1$ indicates salient audio-visual alignment (e.g., an on-screen face visibly speaking in dialogue scenes) and $\synchronyfunc{} = 0$ otherwise (Figure~\ref{fig:av_synchrony}); see Appendix~\ref{sec:vlm_preprocessing} for VLM prompts and more details.
\nchange{All VLM queries use Gemini 3.0 Flash. VLMs are used to annotate source clips during preprocessing (e.g., subject identity and shot type) and to search for cut-on-action candidates, where our formalization bounds the search to the editing window $\EditWindowV{}$ (\Sec{\ref{sec:view}}).}

\section{Evaluation}
\label{sec:evaluation}

\nchange{We first validated our synchrony analysis through a crowdsourced study, and then evaluated \emph{MicroEditor} in} two comparative studies with experienced editors, focusing on our design goals: whether our tools enable more efficient exploration of the \microedit{} space (\DesignGoal{dg:exploration}), and whether our defaults and suggestions help editors identify desirable edits (\DesignGoal{dg:suggestions}). Study~1 (N\,=\,6) evaluated the initial system and validated tools for \rhythm{} and \view{} editing. Based on feedback, we refined the interface and added a Reaction Shot Browser. Study~2 (N\,=\,10) evaluated this expanded system with an additional focus on reaction shot selection and perceived improvements in scene pacing and visual flow. Because micro-editing has no single correct solution and desirable edits depend on creative goals (Section~\ref{sec:design_goals}), our evaluation emphasizes workflow support, decision support, and exploration rather than correctness against a fixed ground truth. Full study protocols and other evaluation details can be found in our supplemental materials.

\begin{table}[h]
\centering
\caption{\nchange{Median (min--max)} ratings (1--7 scale) for Rhythm, Character Focus, Cut-on-Action, Blink, Reaction. Creativity Support: Exploration, Expression, Effort, Immersion, Enjoyment, Efficiency~\cite{cherry2014quantifying}.}
\label{tab:ratings-summary}
\small
\setlength{\tabcolsep}{1.8pt}
\begin{tabular}{lccccc|cccccc}
\hline
& \multicolumn{5}{c|}{\textbf{Features}} & \multicolumn{6}{c}{\textbf{Creativity Support}} \\
& Rhy. & Char. & Act. & Blk. & Rct. & Expl. & Expr. & Effort & Imm. & Enj. & Effi. \\
\hline
Study 1 & 6.0 & \nchange{7.0} & 7.0 & 5.5 & --- & 6.0 & 6.0 & 6.0 & 5.5 & 7.0 & 7.0 \\
& \nchange{\scriptsize 6--7} & \nchange{\scriptsize 6--7} & \nchange{\scriptsize 5--7} & \nchange{\scriptsize 4--7} & --- & \nchange{\scriptsize 5--7} & \nchange{\scriptsize 5--6} & \nchange{\scriptsize 4--6} & \nchange{\scriptsize 5--7} & \nchange{\scriptsize 6--7} & \nchange{\scriptsize 5--7} \\
Study 2 & 6.0 & 6.0 & 6.5 & 4.5 & 7.0 & 6.0 & 5.0 & 6.0 & 5.5 & 7.0 & 7.0 \\
& \nchange{\scriptsize 5--7} & \nchange{\scriptsize 5--7} & \nchange{\scriptsize 4--7} & \nchange{\scriptsize 1--5} & \nchange{\scriptsize 6--7} & \nchange{\scriptsize 5--7} & \nchange{\scriptsize 4--7} & \nchange{\scriptsize 4--7} & \nchange{\scriptsize 3--7} & \nchange{\scriptsize 5--7} & \nchange{\scriptsize 5--7} \\
\hline
\end{tabular}
\end{table}

\subsection{Synchrony Evaluation}
\begin{figure}[t]
  \centering
  \includegraphics[width=\linewidth]{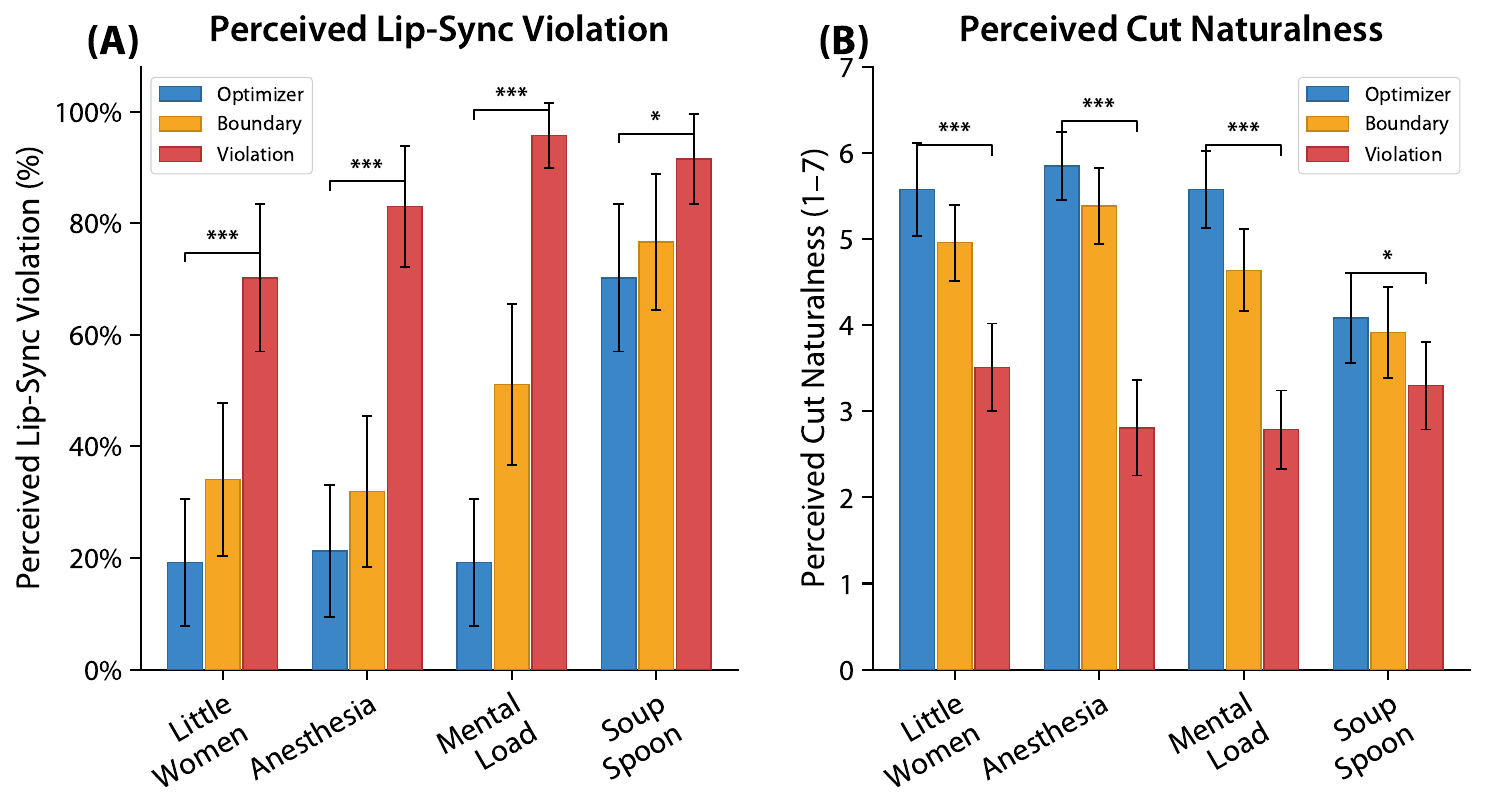}
    \caption{Perceived synchrony ($N=47$) across four scenes. (A)~Perceived lip-sync violation rate (\% responses reporting violation), and (B)~perceived cut naturalness (1--7). Error bars show 95\% CI. Brackets indicate Optimized vs.\ Out-of-Window comparisons within each scene ({*}\,$p<.05$, {***}\,$p<.001$).}
  \Description{Two side-by-side bar charts comparing perceived lip-sync violation and cut naturalness across four video scenes. Both charts group data by three conditions (Optimizer, Boundary, and Violation) for scenes titled Little Women, Anesthesia, Mental Load, and Soup Spoon. Chart A shows that the Optimizer condition consistently yields the lowest lip-sync violation rates (roughly 20 percent for three of the scenes) compared to the Violation condition (which ranges from 70 to 95 percent). Chart B shows the inverse trend for cut naturalness on a 1 to 7 scale: the Optimizer condition consistently scores the highest (averaging around 5.5 for most scenes), while the Violation condition scores the lowest. Brackets confirm statistically significant differences between the Optimizer and Violation conditions across all scenes.}
  \label{fig:perception}
\end{figure}

To validate our method for decoupling audio and video, we conducted a Prolific study ($N=47$; 4 excluded for failing attention checks). Participants viewed 12 randomized clips (${\sim}5$\,s each) across three conditions: Optimized (cut inside $\EditWindow{}$), Boundary (cut on the window edge), and Out-of-Window (fixed offset outside the window). For each clip, participants reported whether they noticed any distracting lip-sync errors or AV mismatches and rated how natural or smooth the cut felt on a 7-point scale (1\,=\,very jarring, 7\,=\,very smooth). Results show a significant main effect of condition on naturalness ($p<.001$, Friedman test). Post-hoc tests confirmed Optimized cuts were rated significantly more natural than both Boundary ($p=.001$) and Out-of-Window ($p<.001$) conditions. In aggregate, Optimized clips achieved the lowest violation rate (32.4\%) and highest naturalness ($M=5.27$), vs.\ Out-of-Window (85.1\%, $M=3.10$) (Figure~\ref{fig:perception}).
Notably, excluding the \emph{Soup Spoon} scene where upstream face-tracking errors produced an unreliable synchrony window, the Optimized violation rate drops to 19.9\% ($M=5.67$). We note that our method's performance is inherently bounded by the capabilities of the underlying AV-sync model.

\subsection{User Studies}
\label{sec:study1}

We recruited 6 professional video editors \nchange{for Study~1} (3 female, 3 male, aged 18--27) with 5--15 years of experience, all primarily using Adobe Premiere Pro\nchange{, and 10 video editors for Study~2 (6 female, 4 male, aged 19--28, 1--10 years of experience) working across narrative films, commercial content, documentaries, and YouTube. Both studies followed the same protocol:} each participant edited two one-minute dialogue scenes (``Pumpkin Princess'' and ``Anesthesia'') under a \textit{manual} condition (all features disabled) and an \textit{assisted} condition (all features enabled), in counterbalanced order.
Sessions lasted 60 minutes.
Participants rated feature usefulness and creativity support on 7-point Likert scales (Table~\ref{tab:ratings-summary}).
\nchange{Study~2 additionally collected ratings on pacing and visual flow, and logged all interactions with millisecond timestamps; one participant's (P9) log was not recorded due to a logging error, so quantitative analyses cover the remaining nine participants ($N=9$).}
Full demographics and per-participant ratings appear in \nchange{Appendices~\ref{sec:userstudy1} and~\ref{sec:userstudy2}; participant IDs below refer to each study separately.}

\namedparagraph{Rhythm Editing}
\nchange{In Study~1, }editors found the pre-roll and trim\slash extend hotkeys useful (median 6.0/7) and efficient (median 7.0/7; \DesignGoal{dg:exploration}).
Pre-roll enabled the repeated listening behavior essential for rhythm editing (\FormFind{ff:rhythm}).
P3 emphasized repetition: ``\textit{As an editor I like to hear and see the cut at least 10 times and it was easy to replay it using the pre-roll key.}''
P6 articulated why context matters: ``\textit{I think the R [pre-roll] feature is something that I really like to just quickly go back... you kind of have to get a little bit of context further, just to get into the flow of that, and then you can watch the cut again.}''

\namedparagraph{Character Focus}
Participants valued the automated L/J cuts as a starting point that offloaded mechanical work for creative decisions.
P3 appreciated the cognitive offloading: ``\textit{You always know in a scene which character is the main one... it was nice to get [character focus] done straight away so I can focus on other things.}''
P2 said: ``\textit{I used one pass to emphasize speaker and was blown away with how accurately it created the space for the character.}''
P1 confirmed: ``\textit{There weren't many specific edits I needed to keep making based off of those automatic cuts, and those cuts would have taken a while manually.}''
Character Focus was rated \nchange{7.0/7} (median), validating \FormFind{ff:views} and \DesignGoal{dg:suggestions}.
\nchange{In Study~2, P3 similarly noted that the automated cuts ``\textit{tended to feel natural and would be similar to what I would do as an editor without these tools}.''}

\namedparagraph{Cut-on-Action}
Cut-on-Action markers (median 7.0/7) transformed the tedious frame-level search into discrete choices (\FormFind{ff:context}).
P5 noted: ``\textit{Cut on Action suggestions took the painstaking process of having to find each moment of large action and match it with another clip and made it extremely simple, with very strong accuracy.}''
P1 described the reduction in decision fatigue: ``\textit{Just having the markers to choose from helped prevent decision fatigue. When you have so many frames versus just the three options, it helps you narrow it down.}''
\nchange{In Study~2, }P4 described using markers for narrative construction: ``\textit{I used this feature to help inform my overall approach to the video narrative for each scene. If there was a particular narrative goal I wanted to achieve, I used the Cut-on-Action as a framework to build it.}''
P3 noted how unexpected suggestions expanded creative thinking: ``\textit{Some of the options were a bit out of the blue even, as those made me realize the amount of flexibility I had in editing, when I would only usually consider an L/J cut difference of only a second.}''
Editing is inherently creative, and there is no fixed ground truth. Still, we observed that when cut-on-action is a strong cue, timelines often show close alignment between manual and \emph{MicroEditor} workflows. For example, the final cut in the \textit{Anesthesia} scene resulted in nearly identical offsets across both methods. Appendix \ref{appendix:timeline-comparisons} provides a visual comparison of these timelines alongside the system's internal synchrony analysis.

\namedparagraph{Blink Markers}
Blink markers were rated 5.5/7 \nchange{in Study~1 and 4.5/7 in Study~2 (medians); they were the only feature to receive a rating below 3 (one editor), and no other feature was rated below 4, as only 4 of 16 editors across both studies reported using cut-before-blink in their regular workflow.}
P3 explained when they matter: ``\textit{When everything is technically right, but it doesn't feel like the cut is right... I use the blink.}''
P5 found them useful for pacing: ``\textit{Using [blink] I was able to keep track of the rhythm of blinking... very helpful in terms of pacing.}''
\nchange{Across the remaining tools, Study~2 ratings were consistent with Study~1, confirming these findings with a larger and more diverse participant pool (Table~\ref{tab:ratings-summary}).}

\begin{figure*}[t]
  \centering
  \includegraphics[width=\textwidth]{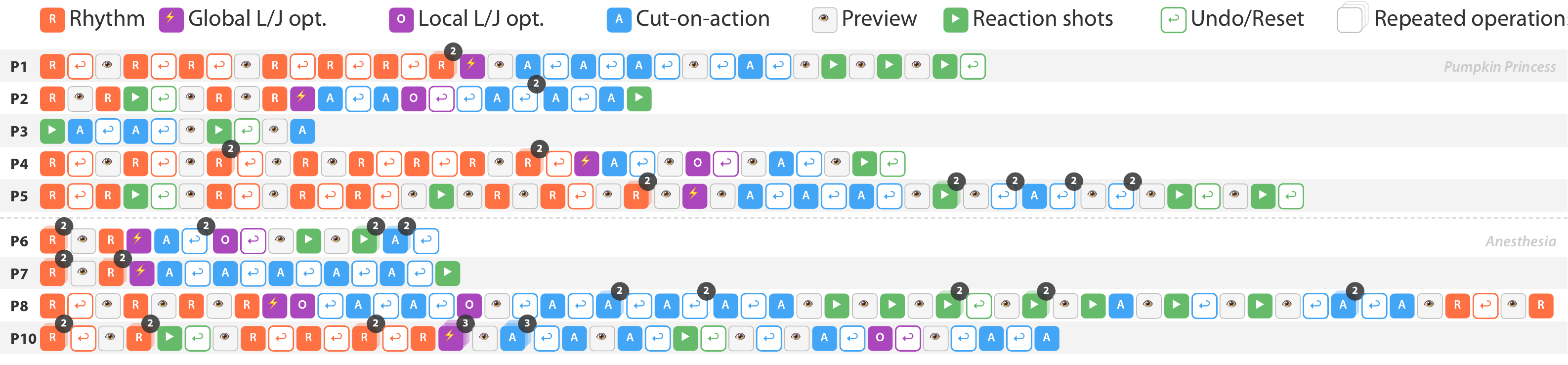}
  \caption{Operation sequences for all nine participants in Study~2. Each icon represents a timeline operation (rhythm edit, marker apply/reset, reaction shot insert, undo, or preview). Stacked icons indicate consecutive repeated operations. \nchange{The top group (P1--P5) edited Pumpkin Princess with MicroEditor; the bottom group (P6--P8, P10) edited Anesthesia with MicroEditor (P9's log was not recorded due to a logging error).}}
  \Description{Sequence charts mapping the chronological editing operations for nine participants. Nine horizontal rows of icons represent specific tool interactions. A clear phased pattern emerges across users: sessions typically begin with "Rhythm" edits to adjust pacing, shift to view edits like "Cut-on-action" and "L/J optimization," and conclude with "Reaction shot" insertions. Frequent "Preview" and "Undo/Reset" icons throughout all phases highlight a highly iterative, trial-and-error editing workflow.}
  \label{fig:operation-sequence}
\end{figure*}

\namedparagraph{Exploration and Iteration}
Participants rated exploration highly (median 6.0/7) and agreed the system encouraged trying different cuts (median 6.5/7).
P5 noted that suggestions led to discoveries: ``\textit{Suggestions came up; I began to think about different edits and ended up choosing several of them.}''
P2 added: ``\textit{Seeing those options does help... Otherwise I would be too rooted in my own interpretation.}''

Participants also requested better support for finding and inserting reaction shots, which required tedious scrubbing through footage (\FormFind{ff:reactions}).
Based on this feedback, we added a Reaction Shot Browser for Study~2.

\namedparagraph{Reaction Shots}
The Reaction Shot Browser was the highest-rated feature (median 7.0/7; 9/10 rated 7/7; \DesignGoal{dg:suggestions}).
Editors described it as ``\textit{Emotionally satisfying}'' (P7), perspective-expanding (P4: ``\textit{I never really used reaction shots as often as I could have, however, this system opened up a perspective}''), and helpful for ``\textit{break[ing] the homogeneity of long dialogues}'' (P5).
Editors used reactions both for masking jump cuts from \rhythm{} tightening (P3, P4, P8, P9) and for directing attention toward the listener's emotional response.
P7 discovered unexpected narrative possibilities: ``\textit{Some of the reaction shots created new meanings (for example, one character's body language slumping in a chair) which influenced the choice of what shot to end the scene with}'' (\DesignGoal{dg:exploration}).
The system supports two complementary insertion methods: highlighting a phrase in the interactive transcript to select semantically meaningful moments, or navigating the playhead for frame-level precision.
P2 noted that ``\textit{the dialogue highlighting feature was also helpful in finding the right time to add a clip.}''

\namedparagraph{Perceived Effort and Output}
Across both studies, participants strongly agreed that \emph{MicroEditor} improved their efficiency (median 7.0/7) and enjoyment (median 7.0/7). Editors also felt the assisted workflow was worth the effort and produced satisfying edits (median 6.0/7 for both). Participants reported better decision support (median 6.0/7 in both studies) and, in Study~2, higher confidence in their editing choices (median 6.0/7 vs.\ 5.0/7 in Study~1). In Study~2, participants felt the assisted edits improved the video output over the rough cut (median 6.0/7). P5 noted ``\textit{the pacing with \nchange{[MicroEditor]} resulted in a much more flowing sequence,}'' and P7 found ``\textit{pacing and flow are better and easier, and save time to iterate and adjust.}'' An interactive gallery of \nchange{video outputs across diverse scenes}, including direct comparisons to the initial rough cuts, is available \nchange{on our project page}.

\begin{table}[t]
  \centering
  \caption{Per-participant interaction metrics in Study~2 ($N=9$, counterbalanced). RS = time in the reaction shot browser. \textbf{Bold} Ops/min indicates the lower value per participant.}
  \label{tab:ops-summary}
  \small
  \setlength{\tabcolsep}{3.6pt}
  \renewcommand{\arraystretch}{1.15}
  \begin{tabular}{l | rrr | rrrrr}
    \hline
    & \multicolumn{3}{c|}{\textbf{Baseline (manual)}} & \multicolumn{5}{c}{\textbf{MicroEditor (assisted)}} \\[2pt]
    \textbf{PID} & Ops & \shortstack[r]{Ops/\compactheaderbreak{}min\,$\downarrow$} & \shortstack[r]{Total\compactheaderbreak{}(min)} & Ops & \shortstack[r]{Ops/\compactheaderbreak{}min\,$\downarrow$} & \shortstack[r]{Edit\compactheaderbreak{}(min)} & \shortstack[r]{RS\compactheaderbreak{}(min)} & \shortstack[r]{Total\compactheaderbreak{}(min)} \\
    \hline
    P1 &  39 & 4.4 &  8.8 & 18 & \textbf{1.3} &  9.2 & 4.7 & 13.9 \\
    P2 &  21 & 2.4 &  8.6 & 17 & \textbf{1.6} &  8.5 & 2.3 & 10.9 \\
    P3 &  37 & 5.0 &  7.4 &  5 & \textbf{1.1} &  3.5 & 1.1 &  4.6 \\
    P4 &  69 & 4.2 & 16.4 & 14 & \textbf{0.9} & 12.5 & 3.4 & 15.9 \\
    P5 &  60 & 10.0 &  6.0 & 25 & \textbf{1.5} & 12.7 & 4.4 & 17.1 \\
    P6 &  51 & 5.3 &  9.5 & 12 & \textbf{1.2} &  6.0 & 3.5 &  9.6 \\
    P7 &  48 & 5.3 &  9.0 & 15 & \textbf{1.6} &  7.3 & 1.4 &  9.5 \\
    P8 &  67 & 3.1 & 21.4 & 40 & \textbf{1.9} & 12.4 & 8.2 & 20.6 \\
    P10 &  64 & 4.1 & 15.8 & 23 & \textbf{1.5} & 12.1 & 3.5 & 15.6 \\
    \hline
    Mean & 50.7 & 4.9 & 11.4 & 18.8 & 1.4 & 9.4 & 3.6 & 13.1 \\
    Median & 51.0 & 4.4 & 9.0 & 17.0 & 1.5 & 9.2 & 3.5 & 13.9 \\
    \hline
  \end{tabular}
\end{table}

\namedparagraph{Quantitative Analysis}
To complement the qualitative findings, we analyzed the interaction logs from Study~2 \nchange{($N=9$; see above)} to understand how the assisted tools changed editing behavior.
Table~\ref{tab:ops-summary} summarizes the key interaction metrics across conditions.

We logged all timeline operations, distinguishing between manual operations (direct clip trims, drags, repositioning) and assisted operations (rhythm edits, applying cut offset suggestions, and inserting reaction shots).
Reaction shot selection uses a modal gallery interface with its own filtering and preview controls, so we report time in this interface separately (RS in Table~\ref{tab:ops-summary}).

Our formative study found that \microediting{} involves tedious frame-by-frame work (\FormFind{ff:everyframe}).
With \emph{MicroEditor}, all nine participants shifted from low-level timeline manipulation to structured tool interactions, with fewer total operations needed to achieve their edits (Table~\ref{tab:ops-summary}).
All nine showed a lower operation rate (mean 1.4 vs.\ 4.9 Ops/min, sign test $p = 0.004$, Cohen's $d = -1.57$), suggesting that the tools compressed what would otherwise require many fine-grained manual adjustments into fewer, higher-level actions.
Although participants perceived the tools as reducing effort (P9: ``\textit{If I were to manually edit, I would be using fewer cuts for awkward silences just because of the manual labor that went into each cut}'', P3: ``\textit{T replaces a lot of clunky movement selecting clips and moving them}''), MicroEditor sessions were not shorter on average (13.1 vs.\ 11.4\,min). Editors appear to have reinvested the reduced mechanical effort into exploring more alternatives, including an average of 3.6 minutes browsing the reaction shot gallery, a modality that does not exist in the baseline condition. Because this additional exploration time makes raw duration comparisons between conditions difficult to interpret, we report operation rates (Ops/min) as the primary measure of mechanical effort.
In the baseline condition, participants were given freeform editing instructions with no requirement to add reaction shots. Despite this, three participants (P6, P7, P8) chose to manually create reaction shots by extending portions of neighboring clips and moving them to the B-roll track. With MicroEditor, the Reaction Shot Browser replaced this manual search entirely. Baseline session times therefore likely undercount the effort needed to search through all available footage for ideal reaction shots. As P2 reflected, without the tools ``\textit{it took a lot of time and I couldn't easily add additional reaction clips.}'' Under time pressure, editors resort to nearby clips instead. P6 noted, ``\textit{Usually I would grab a reaction right before and after if I'm in a time crunch.}''

\namedparagraph{Exploration Patterns}
Beyond aggregate counts, the interaction logs reveal how participants used the assisted tools.
Following prior analyses of creative tool usage patterns~\cite{textoshop2025}, Figure~\ref{fig:operation-sequence} illustrates the full operation sequence for each participant.
Eight of nine participants began their sessions with \rhythm{} edits (\FormFind{ff:rhythm}), adjusting pacing by feel before moving to cut offset suggestions (\FormFind{ff:views}) and reaction shots (\FormFind{ff:reactions}). This ordering is consistent with our formative study, where editors described addressing rhythm before visual attention. P3 was the sole exception, skipping rhythm edits entirely, likely because the rough cut's existing pacing already suited their creative intent.
Seven of nine participants also exhibited iterative refinement at individual cut points, applying a cut offset suggestion, previewing the result in context, resetting, and trying a different suggestion before committing.
Three participants (P1, P2, P8) compared across rule types at the same cut, for example applying a cut-on-action suggestion, resetting, then trying a character-focus suggestion, evaluating how different editorial rationales felt in practice.
Usage patterns varied widely. P3 completed the entire session with only 5 assisted operations (3 cut offset suggestions applied, 2 reaction shots inserted), relying entirely on the system's defaults. P8, the most active explorer, touched every cut point, compared multiple rule types, and inserted 9 reaction shots across 40 total operations.
Similar iterative apply-preview-reset patterns have been observed across other creative domains~\cite{fuzzylinkography2025}.

\section{Discussion}
\label{sec:discussion}

\nchange{Our evaluations focused on how well \emph{MicroEditor} supports our design goals. In this section, we discuss transferable insights from our design (Section~\ref{sec:disc-substrate}), how our tools changed the way editors explored (Section~\ref{sec:disc-exploration}), and limitations and future work (Section~\ref{sec:limitations}).}

\subsection{Discretizing Continuous Creative Spaces}
\label{sec:disc-substrate}

\begin{nhan_revised}
The space of possible \microedits{} is continuous: audio and video cuts can be decoupled by any offset within $\EditWindow{}{}$. It is also multidimensional, since each cut involves competing objectives like rhythm, character emphasis, and action continuity. Editors reason about this space through discrete heuristics instead: cut-on-action, cut-before-blink, favor a character's screen time. Our core contribution is turning these heuristics into navigable structures built on audiovisual synchrony windows, which serve two purposes at once. For the editor, they guarantee perceptual safety: any offset within $\EditWindow{}{v}$ preserves audiovisual flow, so the editor can focus on creative intent instead of watching for lip-sync errors. For the system, they shrink the AI's search space, limiting the frames a VLM must reason about to a region where every candidate is already valid. Recent work calls this kind of shared structure a \emph{compositional substrate}~\cite{cao2025compositional}. It makes both the human's and the AI's work interpretable to each other. Our system goes further by populating these windows with categorized alternatives, so they become a structure for composition, not a boundary that excludes bad options. The same idea applies to reaction shots, where low-synchrony regions become candidates the system indexes by emotion and shot size for browsing (\FormFind{ff:reactions}).
\end{nhan_revised}

\begin{nhan_revised}
More broadly, this suggests three design principles for creativity support in continuous perceptual parameter spaces, such as photo and video color grading or character animation:
\begin{packed_item}
\item Formalize domain constraints to compute boundaries within which any choice is perceptually safe.
\item Within those boundaries, surface alternatives categorized by domain-specific rationale rather than ranked by a single score.
\item A single computed constraint can serve more than one purpose: the same boundary that defines when one operation is safe can also define candidates for a complementary operation.
\end{packed_item}
We demonstrate these principles in film \microediting{}, but they should apply more broadly. In AI-assisted photo color grading, for instance, perceptual constraints (skin tones stay natural, highlights do not clip) could define safe boundaries within which generated alternatives are grouped by aesthetic intent rather than ranked by a single quality score. In character animation, the same skeletal rig that keeps a motion perceptually plausible also lets an animator reuse it, like walking, across many different characters. Applying a motion template like this, instead of tuning each joint's rotation knobs one by one, turns a continuous space of poses into multiple, easy-to-navigate discrete choices.
\end{nhan_revised}

\subsection{Lowering the Cost of Exploration}
\label{sec:disc-exploration}

\nchange{Because any point within $\EditWindow{}$ is perceptually safe, the system can surface multiple discrete suggestions at each cut, categorized by editorial rationale, and populate a reaction shot browser from low-synchrony regions in the source footage. This categorization is a deliberate design choice. A single ``best'' AI output often induces design fixation~\cite{wadinambiarachchi2024}, and when AI auto-fills an entire structure, the perceived cost of revision stifles exploration~\cite{cao2025compositional}. To avoid premature convergence,} our approach explicitly extracts \emph{dimensions} (editorial rationales) across which users can explore alternatives, making the continuous design space legible rather than merely navigable~\cite{luminate2024}.

\nchange{The interaction logs suggest this strategy worked: participants compared suggestions drawn from different rationales at the same cut point before committing (Figure~\ref{fig:operation-sequence}), a rapid comparison that depends on alternatives being computationally and interactively cheap~\cite{huh2025videodiff}.} By never auto-applying suggestions, our system lets editors try with one click and revert with another (\DesignGoal{dg:suggestions}).

The quantitative logs tell the same story at a broader scale. All nine logged participants showed a lower operation rate with \emph{MicroEditor} than in the baseline \nchange{(Table~\ref{tab:ops-summary})}, a shift from frame-level adjustments to higher-level creative evaluations.
\begin{nhan_revised}
Lowering the cost per attempt did not make sessions shorter. Instead, editors spent the saved mechanical effort on trying more alternatives, an instance of what \citet{olsen2007} calls \textit{expressive leverage}. When a tool is effective, it does more than speed up standard tasks. It changes what creators choose to do~\cite{shneiderman2006cst}.
\end{nhan_revised}
P6 explained: ``\textit{when you're on a time crunch, you don't care that much... you just go with the first thing you see that works. But here... the cost of trying is a lot lower}'' (\DesignGoal{dg:exploration}).

The tedious part of \microediting{} is finding viable candidates, not choosing between them.
\begin{nhan_revised}
\citet{dmytryk2018film} makes the same distinction: editors settle for simpler cuts when better ones take ``more time and greater effort,'' yet the cut itself depends on a sense of timing that defines exceptional editors. We see the same pattern here: once search became cheap, editors used techniques they already knew more often, but still made every timing decision themselves. This points to where automation belongs in creative tools: the search that comes before a decision.
\end{nhan_revised}
Automating the search while preserving the choice respects the expertise that editors bring to every cut.

\subsection{Limitations and Future Work}
\label{sec:limitations}

\namedparagraph{Model Dependencies}
Our system relies on off-the-shelf models for synchrony estimation (LR-ASD, AVID-CMA), blink detection (MediaPipe), and a VLM (Gemini 3.0 Flash) for coarse scene analysis and cut-on-action suggestions via numbered mosaics, and is bounded by their current capabilities. \nchange{Current VLMs lack fine-grained temporal understanding~\cite{qu2025does}. Synchrony models can also struggle with multiple simultaneous speakers or overlapping dialogue, and we assume a single mixed audio track. These models are pluggable components in our formalization rather than contributions of their own, so improvements to any of them would improve the system directly, without changing the interaction design.}

\namedparagraph{Scope and Generalization}
\nchange{Our implementation and evaluation focus on dialogue scenes. Scene types where \synchrony{} plays a smaller role (voiceovers, music-driven scenes, or action scenes without dialogue) would require new definitions of synchrony and rhythm, which we leave to future work. The value of \microediting{} also depends on the starting rough cut, and there is no single ground-truth final cut: editing styles varied widely even among our participants (Figure~\ref{fig:operation-sequence}, Table~\ref{tab:ops-summary}). Finally, our formalization treats each \microedit{} as a local decision, but editors may want consistency across cuts, such as cutting consistently early or late on speaker changes. Conditioning suggestions on an editor's prior decisions is another direction for future work.}

\namedparagraph{Informing Agentic and Generative Video Tools} Current generative video models produce tightly synchronized audio-video outputs with no mechanism for the audiovisual decoupling that professional editors rely on to control pacing and attention. Our formalization of the \microedit{} space provides a structure for future AI systems. By applying these synchrony bounds, both agentic editing tools and video models could produce outputs with explicit rhythm and view control, bridging \microediting{} and \macroediting{}.

\section{Conclusion}
\label{sec:conclusion}

While rough cuts establish a film’s narrative, the transition to a polished fine cut requires navigating a continuous, multidimensional space of \textit{\microedits{}} that has previously remained informal and manual. Our work formalizes this stage by decoupling audio and video through \textit{synchrony windows}, transforming a tedious, frame-by-frame process into a structured computational space. By surfacing specific micro-edit types, from rhythm control to viewer attention (cut-on-action, cut-before-blink, and reaction-shot alternatives), we provide a representation that bridges the gap between raw timeline operations and the editor's intent.
Editors in our studies reported that this structure reduced mechanical effort and allowed them to reinvest their focus into higher-level creative goals. 
More broadly, this work demonstrates that computational assistance in creative workflows is most effective when it operates at this intermediate granularity: providing the structure to navigate complex decision spaces without automating the creative outcome.

\begin{acks}
We thank Mira Dontcheva and Fabian Caba Heilbron at Adobe Research for valuable discussions during the course of this work. We are especially grateful to the professional editors who participated in our formative study and user studies for generously sharing their time, candor, and insight into their craft.
\end{acks}

\bibliographystyle{ACM-Reference-Format}
\bibliography{bibliography}

\clearpage

\clearpage
\appendix

\section{Qualitative Timeline Comparisons}
\label{appendix:timeline-comparisons}

\begin{figure}[b]
  \centering
  \includegraphics[width=\columnwidth]{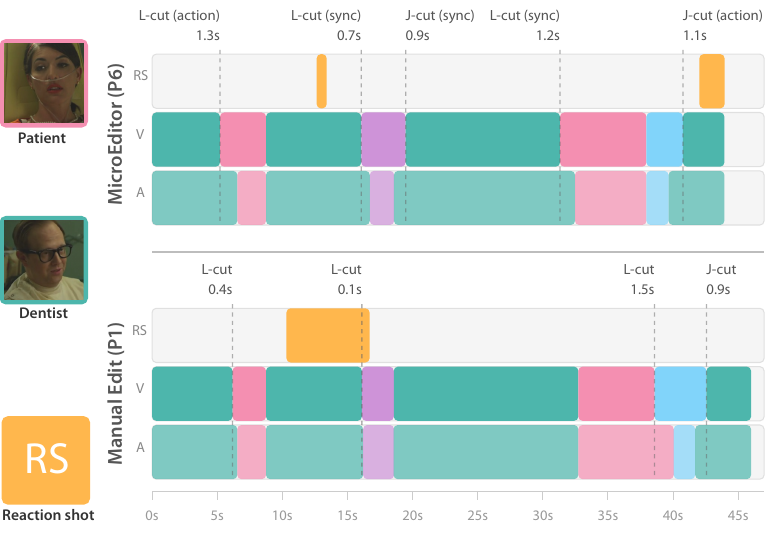}
  \caption{Two participants editing the same scene \textit{Anesthesia}: P6 with \nchange{\emph{MicroEditor}} (top) and P1 editing manually (bottom).
  The \nchange{\emph{MicroEditor}} row labels cut types per user's decisions (e.g., optimization based on AV synchrony, cut-on-action). When cut-on-action is a strong cue, manual and assisted edits sometimes converge on similar cut timing (0.9 vs 1.1s J-cut offset). Video frames~\textcopyright~EditStock.}
  \Description{Two timeline diagrams of the same scene, one edited with MicroEditor by participant P6 and one edited manually by participant P1. Each shows reaction shot, video, and audio tracks with colored segments per speaker and dashed lines marking cuts, labeled with L-cut and J-cut offsets in seconds. Thumbnails of the two characters and a reaction shot icon appear on the left.}
  \label{fig:timeline-comparison}
\end{figure}

\begin{figure*}[h]
    \centering
    \includegraphics[width=\textwidth]{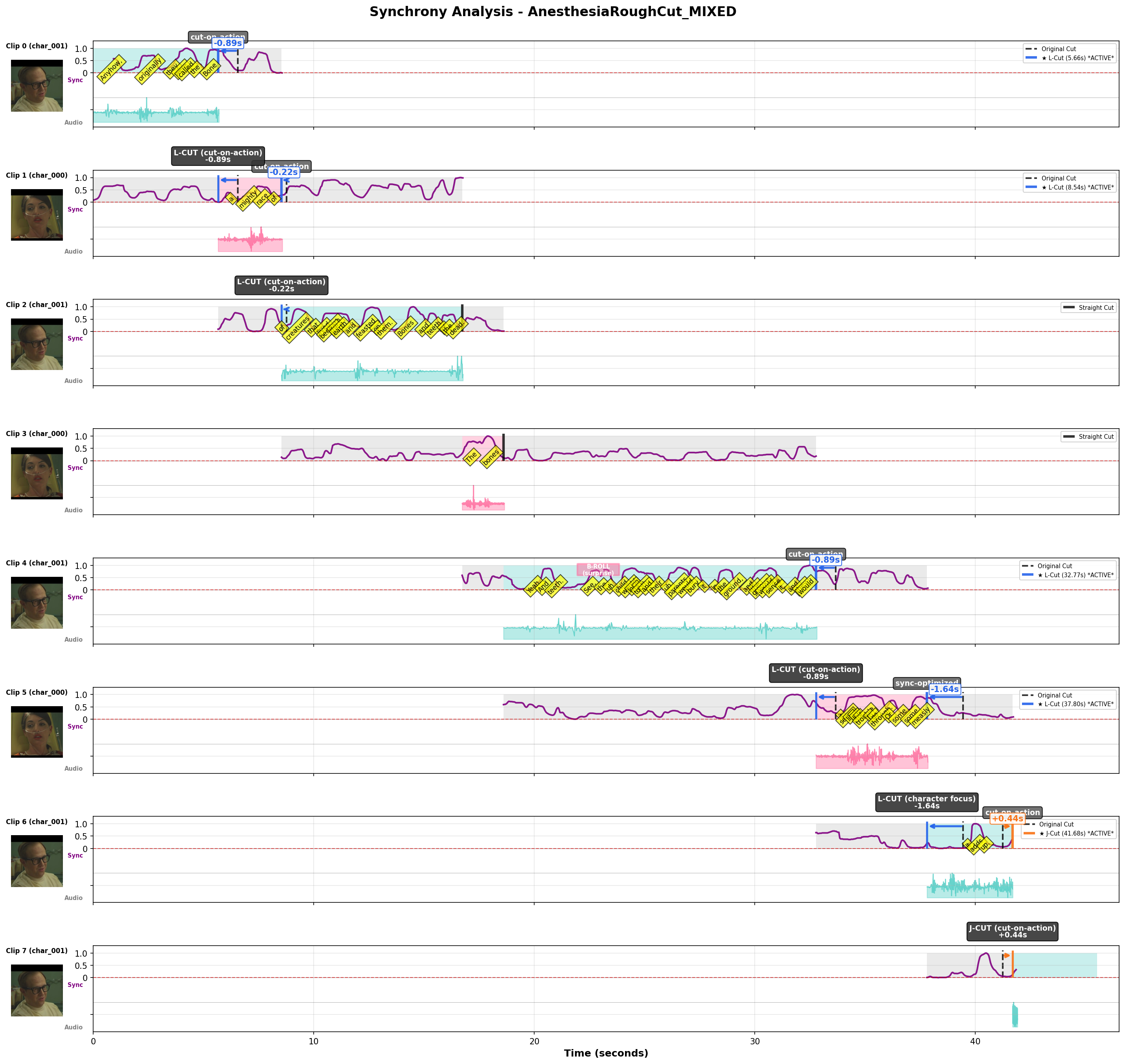} 
    \caption{System-level synchrony analysis for an edit of \textit{Anesthesia} scene. The visualization shows the continuous synchrony signal ($\sigma$, purple line), word-level timing (yellow boxes), and the calculated safe decoupling windows used to generate the L/J-cut and Cut-on-Action suggestions surfaced to the user. Video frames~\textcopyright~EditStock.}
    \Description{Eight stacked plots, one per clip of the scene, each showing a continuous synchrony curve, word-level speech boxes, the audio waveform, and labeled markers for the applied cut type and offset. A small character thumbnail identifies each clip at the left.}
    \label{fig:synchrony_analysis}
\end{figure*}

This appendix visualizes the final timelines for the \textit{Anesthesia} scene to compare the baseline manual workflow (P1) with \emph{MicroEditor} (P6). As noted in Section 6.2, micro-editing is subjective. Overall pacing and reaction shot placements naturally vary based on individual editing styles.
However, when perceptual cues are strong, we observe similar timing outcomes. Figure~\ref{fig:timeline-comparison} shows both participants arriving at a comparable J-cut offset (0.9s vs.\ 1.1s) near the end of the scene, driven by a cut-on-action. 
To illustrate the underlying method that generates these suggestions, Figure~\ref{fig:synchrony_analysis} provides a detailed view of the system's internal synchrony model for the same \textit{Anesthesia} scene. It visualizes the word-level ASR alignments, the continuous audio-visual synchrony curves ($\sigma$), and the resulting calculated bounds that constrain the suggestions surfaced in UI. 

Similar timeline visualizations and synchrony analyses for other scenes can be found in \nchange{the interactive gallery on our project page}.

\section{Preprocessing Pipeline}
\label{sec:vlm_preprocessing}

This section details the offline preprocessing that extracts the signals used by the interactive tools described in \Sec{\ref{sec:system_design}}. Given source footage $\SourceMediaSet{}$ and a rough cut $\RoughCut{}$, preprocessing runs in three stages before the editing session begins (Figure~\ref{fig:system_overview}).

\subsection{System Implementation and Prompt Details}
Since VLMs excel at coarse scene understanding but struggle with micro-temporal reasoning due to their inherent sparse sampling~\cite{qu2025does}, we use VLMs for semantic analysis (character identification, shot classification) while relying on traditional computer vision for frame-level signals (synchrony proxy, blink detection).

\begin{figure}
      \centering
      \includegraphics[width=\columnwidth]{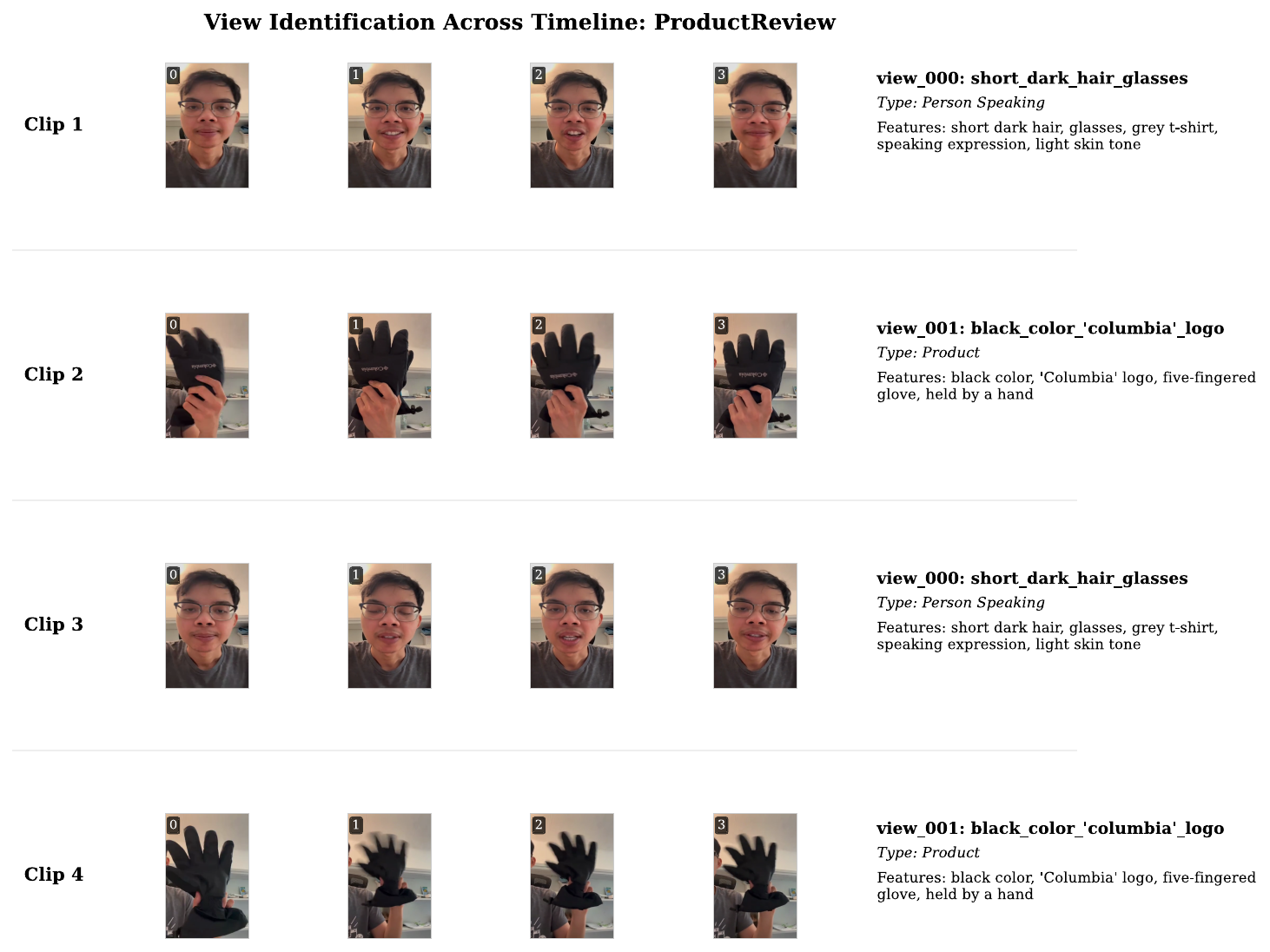}
      \caption{
          View identification using vision-language model analysis, shown on a product review scene captured by the authors.
          The VLM distinguishes the human presenter from product demonstration shots and extracts distinctive features for each view, enabling the Character Focus feature to redistribute screen time through automated L/J cut suggestions.
      }
      \Description{Four rows of video frames from a product review. Rows alternate between a presenter speaking to the camera and close-ups of a glove held up to the camera. Beside each row, the vision-language model's view label and extracted visual features are listed.}
      \label{fig:view_identification}
  \end{figure}

\namedparagraph{Stage 1: VLM Coarse Analysis (2--5 min)}
A VLM (Gemini) analyzes all source footage sampled at 1~FPS to within 1 API call to cross-associate semantic information and produce a \emph{global character map}. For each source clip $\MediaMi{j} \in \SourceMediaSet{}$, the VLM identifies distinct characters, shot types, and non-usable regions. We use 480p proxy files to reduce token cost while preserving sufficient detail for character recognition. The prompt structure:

\begin{quote}
\small
\textit{Analyze ALL footage files together to identify characters and shot types.}

\textbf{CHARACTER IDENTIFICATION:} \textit{Identify all distinct people and objects of interest across all footage files (Figure~\ref{fig:view_identification}). The same person in different footage = same character\_id. Look for: hair, clothing, glasses, facial features, body type. A close-up in one footage and wide shot in another can be the same person.}

\textbf{SHOT TYPE CLASSIFICATION:} \textit{Classify each segment as: extreme\_close\_up, close\_up, medium\_close\_up, medium, medium\_wide, wide, over\_the\_shoulder, two\_shot, or group\_shot.}

\textbf{OVER-THE-SHOULDER RULE:} \textit{In OTS shots, assign character\_id to the person whose face is visible (not the person shown from behind). This ensures visible faces for reaction shot extraction.}

\textbf{NON-USABLE SEGMENTS:} \textit{Mark slates, crew visible, camera movement, and empty frames as no\_character.}

\textbf{TEMPORAL COVERAGE:} \textit{Cover the entire duration of each footage file with no gaps. Approximate timing ($\pm$1--2 seconds) is acceptable.}
\end{quote}

\namedparagraph{Stage 2: AV Sync Processing (2--5 min)}
In parallel, we extract frame-level signals locally:
\begin{packed_item}
\item \textbf{Word boundaries}: ASR (Parakeet-TDT~\cite{sekoyan2025canary} or Whisper~\cite{radford2023robust}) provides word-level timestamps, defining bounds for \rhythm{} editing windows.
\item \textbf{Synchrony function}: For dialogue scenes, active speaker detection (LR-ASD~\cite{liao2025lr}) computes per-frame lip-audio correlation for each face track. For non-dialogue content (e.g., sports, music), we use an audio-visual alignment network (AVID-CMA~\cite{morgado2021audio}).  We threshold the continuous signal to obtain $\synchronyfunc{}(\MediaMi{j}, t) \in \{0, 1\}$.
\item \textbf{Blink onsets}: MediaPipe~\cite{lugaresi2019mediapipe} estimates eye aspect ratio per frame; blink onset occurs when EAR falls below threshold.
\end{packed_item}

\namedparagraph{Stage 3: VLM-ASD Correspondence (instant)}
We link VLM character identities to ASD face tracks by extracting representative face crops from each track and querying Gemini to match them against VLM-generated character thumbnails. When matching is ambiguous, we fall back to DeepFace embeddings (Facenet512) with cosine similarity. This enables per-character queries of $\synchronyfunc{}$ and supports the character focus optimization (\Sec{\ref{sec:view}}).

\subsection{Cut-on-Action Detection}

Current state of VLMs cannot reliably reason about frame-level motion continuity~\cite{qu2025does}. We address this by converting the temporal reasoning problem into a spatial matching task. The model is prompted with the 2$\times$2 grid layout (sample from the left and right bounds of $W$) and temporal alignment rules:

\begin{quote}
\small
\textit{You are a professional film editor with expertise in cut-on-action techniques.}

\textbf{GRID LAYOUT:} \textit{Top row shows earlier cut opportunities (outgoing end, incoming preroll). Bottom row shows later cut opportunities (outgoing post-roll, incoming start).}

\textbf{TEMPORAL ALIGNMENT RULES:}
\begin{itemize}
    \item \textit{For later cuts (J-cut): Match frames with equal numbers (A=B), since frame 1 = cut point for both grids.}
    \item \textit{For earlier cuts (L-cut): Match frames by distance from last frame, maintaining consistent offset (B$-$A).}
\end{itemize}

\textbf{FOCUS ON:} \textit{Motion trajectory, direction continuity, body alignment (head angle, shoulder rotation, limb positions), and temporal alignment.}
\end{quote}

The temporal alignment rules encode the key insight that frames are sampled at the same rate (4.5\,fps), so valid cut points must match frames representing the same moment relative to the cut boundary.

\namedparagraph{Motion Matching}
The VLM identifies frame pairs where motion trajectory, body position, or gaze continues smoothly across the cut boundary. Each pair $(n_\text{out}, n_\text{in})$ converts to a timeline offset:
\[
\delta = (n_{\text{frame}} - 1) / \text{fps}_{\text{mosaic}}
\]
Top-row matches yield L-cuts ($\delta < 0$). Bottom-row matches yield J-cuts ($\delta > 0$).

\subsection{Reaction Shot Indexing}

Low-sync segments provide source material for reaction shots. A segment from $\MediaMi{j}$ qualifies as a candidate if:
\begin{packed_item}
\item $\synchronyfunc{}(\MediaMi{j}, t) = 0$ throughout (character visible but not speaking)
\item The target character is detected in $\geq$80\% of frames
\item The segment is adjacent to spoken material (filtering irrelevant silent stretches)
\end{packed_item}
We index candidates by character identity and emotion embeddings (EmoCLIP~\cite{foteinopoulou2024emoclip}), enabling the semantic search interface described in \Sec{\ref{sec:reactionshots}}.

\subsection{Processing Performance}

Table~\ref{tab:preprocessing_time} shows typical times for a one-minute dialogue scene with 4 source clips ($\sim$5 minutes raw footage).

\begin{table}[h]
\centering
\caption{Preprocessing time breakdown (1-minute scene, 4 source clips, MacBook Pro M2)}
\label{tab:preprocessing_time}
\small
\begin{tabular}{l|r}
\hline
\textbf{Stage / Component} & \textbf{Time} \\
\hline
\textit{Stage 1: VLM Coarse Analysis} & \\
\quad Proxy generation (480p) & 30s \\
\quad VLM footage analysis (Gemini) & 90s \\
\hline
\textit{Stage 2: AV Sync Processing} & \\
\quad ASR transcription & 45s \\
\quad Active speaker detection (LR-ASD) & 60s \\
\quad Blink detection (MediaPipe) & 15s \\
\hline
\textit{Stage 3: VLM-ASD Matching} & \\
\quad Face embedding + matching & 20s \\
\hline
\textit{Per-Cut Processing} & \\
\quad Cut-on-action mosaics + VLM & 60s \\
\hline
\textbf{Total} & $\sim$5 min \\
\hline
\end{tabular}
\end{table}

\noindent Stages 1 and 2 run in parallel; VLM API calls are the primary bottleneck. All signals are computed in source time and remain stable across timeline changes, so preprocessing runs once per project.

\section{ User Study 1 Details}
\label{sec:userstudy1}

Building on the insights identified in our formative study (Section \ref{sec:formative_concise}), we evaluated how \emph{MicroEditor} supports (1) \textbf{rhythm} (temporal flow) and (2) \textbf{view} (visual continuity and narrative focus) while reducing cognitive load and encouraging creative iteration.

\subsection{Participants}

We recruited 6 professional video editors (3 women, 3 men, aged 18-27) with 5-15 years of editing experience (Table \ref{tab:participants}). One edits daily (P3), three edit weekly or multiple times per week (P1, P6 edit weekly; P3 daily), two edit monthly (P2, P5), and one edits yearly (P4). Participants work across narrative films (5 editors), documentaries (3 editors), YouTube content (2 editors), and social media content (2 editors). All primarily use Adobe Premiere Pro in their professional work.

\begin{table*}
\centering
\caption{Participant demographics and feature ratings from the MicroEditor user study}
\label{tab:participants}
\small
\begin{tabular}{c|c|c|l|l||c|c|c|c}
\hline
\multirow{2}{*}{\textbf{ID}} & \multicolumn{4}{c||}{\textbf{Demographics}} & \multicolumn{4}{c}{\textbf{MicroEditor Feature Ratings}} \\
\cline{2-9}
& \textbf{Age} & \textbf{Gender} & \textbf{Experience} & \textbf{Content Types} & \textbf{Rhythm} & \textbf{Character Focus} & \textbf{Action} & \textbf{Blink} \\
\hline
P1 & 22 & F & 7+ years, Premiere & Narrative, YouTube, TikTok & 6 & 7 & 5 & 7 \\
\hline
P2 & 27 & M & 15 years & Narrative, Documentary & 6 & 6 & 7 & 5 \\
\hline
P3 & 26 & F & 7 years, Premiere & Narrative, Documentary & 7 & 7 & 6 & 4 \\
\hline
P4 & 26 & F & 5 years & Narrative, YouTube & 6 & 7 & 7 & 4 \\
\hline
P5 & 18 & M & 6 years & Narrative, Digital Media & 6 & 7 & 7 & 6 \\
\hline
P6 & 19 & M & 8 years & Narrative, Documentary & 6 & 6 & 7 & 7 \\
\hline
\end{tabular}
\vspace{2mm}

\footnotesize
\textbf{Note:} Participants completed two editing tasks using the \emph{MicroEditor} interface in counterbalanced order: manual editing (all features disabled) and assisted editing (features enabled). Scene assignment (Pumpkin Princess/Anesthesia) and condition order were counterbalanced. Feature ratings on 1-7 scale where 7 = Extremely Useful. Rhythm = Pre-roll \& Trim, Character Focus = Emphasize Speaker (automatic L/J cuts), Action = Cut-on-action suggestions, Blink = Blink detection markers.
\end{table*}

\subsection{Study Format}

All participants completed a 60-minute in-person session. After a tutorial on \emph{MicroEditor} terminology and features, participants edited two 1-minute dialogue scenes using our interface: ``Pumpkin Princess'' (awkward romantic interaction between a nerdy outcast and her crush) and ``Anesthesia'' (suspenseful 1950s scene where a housewife suspects her dentist). Participants completed both conditions in counterbalanced order: manual editing (all features disabled) and assisted editing (features enabled). We encouraged think-aloud protocol throughout. Following each edit, participants rated features on 7-point Likert scales and completed creativity support metrics adapted from the Creativity Support Index (Table \ref{tab:csi-experience}).

\begin{table*}
\centering
\caption{Creativity support and user experience metrics. Participants rated agreement on a 7-point scale (1=Strongly Disagree, 7=Strongly Agree) with statements adapted from the Creativity Support Index and task-specific measures.}
\label{tab:csi-experience}
\small
\begin{tabular}{l|c|c|c|c|c|c||c}
\hline
\textbf{Metric} & \textbf{P1} & \textbf{P2} & \textbf{P3} & \textbf{P4} & \textbf{P5} & \textbf{P6} & \textbf{Median} \\
\hline
\multicolumn{8}{l}{\textit{Creativity Support Dimensions}} \\
\hline
Exploration: Helped explore possibilities & 6 & 7 & 6 & 6 & 6 & 5 & 6.0 \\
\hline
Exploration: Encouraged different cuts & 7 & 7 & 7 & 6 & 6 & 4 & 6.5 \\
\hline
Expressiveness: Achieved creative vision & 5 & 5 & 6 & 6 & 6 & 6 & 6.0 \\
\hline
Expressiveness: Achieved editing style & 6 & 5 & 6 & 6 & 5 & 5 & 5.5 \\
\hline
Effort/Reward: Quality worth effort & 6 & 6 & 6 & 6 & 6 & 4 & 6.0 \\
\hline
Results Worth Effort: Satisfied with output & 5 & 5 & 6 & 6 & 6 & 7 & 6.0 \\
\hline
Immersion: Attention on task not tool & 5 & 5 & 6 & 5 & 7 & 6 & 5.5 \\
\hline
Enjoyment: Enjoyed using system & 6 & 7 & 7 & 7 & 7 & 6 & 7.0 \\
\hline
\multicolumn{8}{l}{\textit{Task-Specific Efficiency}} \\
\hline
Decision Support: Better editing decisions & 6 & 7 & 7 & 6 & 6 & 5 & 6.0 \\
\hline
Confidence: More confident in choices & 6 & 5 & 5 & 6 & 5 & 4 & 5.0 \\
\hline
Efficiency: Saved time & 5 & 7 & 7 & 7 & 7 & 6 & 7.0 \\
\hline
\end{tabular}
\end{table*}

\section{Full User Study 2 Details}
\label{sec:userstudy2}

Building on User Study 1 (Sec \ref{sec:userstudy1}), we conducted a second evaluation to assess the newly added \textbf{Reaction Shot Selection} feature with emotion filtering, while continuing to evaluate the core rhythm and view pass capabilities with a larger participant pool.

\subsection{Participants}

We recruited 10 video editors (6 women, 4 men, aged 19-28) with 1-10 years of editing experience (Table \ref{tab:participants2}). Participants edit at varying frequencies: two edit weekly (P2, P8), three edit monthly (P4, P7, P9), four edit every few weeks to monthly (P1, P3, P5, P6), and one edits a few times per year (P10). Participants work across narrative films (7 editors), commercial/social media content (5 editors), documentaries (3 editors), and YouTube/digital content (3 editors). Most primarily use Adobe Premiere Pro, with some using DaVinci Resolve or Final Cut Pro.

\begin{table*}
\centering
\caption{Participant demographics and feature ratings from User Study 2 (with Reaction Shot Retrieval)}
\label{tab:participants2}
\small
\begin{tabular}{c|c|c|l|l||c|c|c|c|c}
\hline
\multirow{2}{*}{\textbf{ID}} & \multicolumn{4}{c||}{\textbf{Demographics}} & \multicolumn{5}{c}{\textbf{MicroEditor Feature Ratings}} \\
\cline{2-10}
& \textbf{Age} & \textbf{Gender} & \textbf{Experience} & \textbf{Content Types} & \textbf{Rhythm} & \textbf{Char. Focus} & \textbf{Action} & \textbf{Blink} & \textbf{Reaction} \\
\hline
P1 & 20 & F & 10+ years & Narrative, Animation & 6 & 6 & 7 & 4 & 6 \\
\hline
P2 & 20 & F & 3 years & Dance, Trailers & 6 & 7 & 6 & 5 & 7 \\
\hline
P3 & 20 & M & 3-4 years & Reels, Product Videos & 5 & 6 & 4 & 1 & 7 \\
\hline
P4 & 24 & M & 7 years & Narrative, News & 7 & 6 & 7 & 5 & 7 \\
\hline
P5 & 19 & F & 4 years & Narrative, Commercial & 6 & 7 & 6 & 3 & 7 \\
\hline
P6 & 27 & F & 3+ years & Narrative, Architectural & 7 & 6 & 7 & 5 & 7 \\
\hline
P7 & 27 & M & 10 years & Narrative & 6 & 5 & 6 & 5 & 7 \\
\hline
P8 & 24 & F & 1 year & Short Reels/TikTok & 7 & 7 & 7 & 4 & 7 \\
\hline
P9 & 25 & F & 3 years & Music Videos, Interviews & 5 & 6 & 5 & 5 & 7 \\
\hline
P10 & 28 & M & 10 years & Narrative, Documentary & 6 & 6 & 7 & 4 & 7 \\
\hline
\end{tabular}
\vspace{2mm}

\footnotesize
\textbf{Note:} Feature ratings on 1-7 scale where 7 = Extremely Useful. Rhythm = Pacing Control via Trim/Extend, Char. Focus = Character Focus with automated L/J cuts, Action = Cut-on-action suggestions, Blink = Blink detection markers, Reaction = Reaction Shot Selection with Emotion Filtering.
\end{table*}

\subsection{Study Format}

All participants completed a 60-minute in-person session following the same protocol as User Study 1. After a tutorial covering \emph{MicroEditor} features including the new Reaction Shot Selection tool, participants edited two 1-minute dialogue scenes (``Pumpkin Princess'' and ``Anesthesia'') in counterbalanced order across manual and assisted conditions. Following each edit, participants rated features on 7-point Likert scales and completed creativity support metrics (Table \ref{tab:csi-experience2}). Additionally, participants rated the degree to which pacing and visual flow improved over the original rough assembly.

\begin{table*}
\centering
\caption{Creativity support and user experience metrics for User Study 2. Participants rated agreement on a 7-point scale (1=Strongly Disagree, 7=Strongly Agree).}
\label{tab:csi-experience2}
\small
\begin{tabular}{l|c|c|c|c|c|c|c|c|c|c||c}
\hline
\textbf{Metric} & \textbf{P1} & \textbf{P2} & \textbf{P3} & \textbf{P4} & \textbf{P5} & \textbf{P6} & \textbf{P7} & \textbf{P8} & \textbf{P9} & \textbf{P10} & \textbf{Med.} \\
\hline
\multicolumn{12}{l}{\textit{Creativity Support Dimensions}} \\
\hline
Exploration: Helped explore possibilities & 6 & 6 & 5 & 7 & 7 & 7 & 6 & 6 & 6 & 6 & 6.0 \\
\hline
Exploration: Encouraged different cuts & 7 & 5 & 6 & 6 & 7 & 7 & 6 & 6 & 6 & 6 & 6.0 \\
\hline
Expressiveness: Achieved creative vision & 5 & 5 & 6 & 4 & 5 & 7 & 5 & 6 & 5 & 6 & 5.0 \\
\hline
Expressiveness: Achieved editing style & 5 & 6 & 6 & 5 & 6 & 7 & 6 & 6 & 5 & 6 & 6.0 \\
\hline
Effort/Reward: Quality worth effort & 6 & 6 & 5 & 4 & 7 & 7 & 6 & 6 & 5 & 7 & 6.0 \\
\hline
Results Worth Effort: Satisfied with output & 6 & 7 & 6 & 6 & 7 & 7 & 6 & 6 & 4 & 7 & 6.0 \\
\hline
Immersion: Attention on task not tool & 5 & 6 & 4 & 7 & 5 & 6 & 7 & 6 & 3 & 5 & 5.5 \\
\hline
Enjoyment: Enjoyed using system & 6 & 7 & 5 & 7 & 7 & 7 & 6 & 7 & 5 & 7 & 7.0 \\
\hline
\multicolumn{12}{l}{\textit{Task-Specific Efficiency}} \\
\hline
Decision Support: Better editing decisions & 6 & 6 & 6 & 6 & 7 & 7 & 6 & 7 & 4 & 6 & 6.0 \\
\hline
Confidence: More confident in choices & 5 & 7 & 6 & 5 & 7 & 7 & 5 & 6 & 5 & 6 & 6.0 \\
\hline
Efficiency: Saved time & 7 & 7 & 6 & 7 & 7 & 7 & 7 & 7 & 5 & 7 & 7.0 \\
\hline
\multicolumn{12}{l}{\textit{Additional Rating}} \\
\hline
Pacing/Flow: Improved over rough assembly & 7 & 6 & 6 & 6 & 6 & 7 & 6 & 6 & 5 & 7 & 6.0 \\
\hline
\end{tabular}
\end{table*}

\section{Synchrony Study Details}
\label{sec:userstudy3}
For each of the 12 video clips, participants on Prolific.com were shown the embedded video player and asked the following two questions. The questions were repeated identically for every clip.

\vspace{1em}
\noindent
\textbf{Q1: Did you notice any distracting audio-visual mismatches or lip-sync errors in this clip?}
\begin{itemize}[label=$\circ$]
    \item Yes
    \item No
\end{itemize}

\vspace{1em}
\noindent
\textbf{Q2: Overall, how natural did the flow of this clip feel?}
\begin{itemize}[label=$\circ$]
    \item 1 - Very jarring / distracting
    \item 2
    \item 3
    \item 4 - Neutral
    \item 5
    \item 6
    \item 7 - Very smooth / natural
\end{itemize}

\end{document}
\endinput